\pdfoutput=1

\documentclass{article}

\usepackage{amsmath}
\usepackage{amssymb}
\usepackage{tikz}
\usepackage{bm}
\usepackage[english]{babel}
\usepackage{graphicx}
\usepackage{microtype,lmodern}

\newcommand{\diff}{\mathrm{d}}
\newcommand{\ad}{\mathrm{Ad}}
\newcommand{\tr}{\mathrm{Tr}}
\newcommand{\pointradius}{1pt}

\renewcommand{\i}{\mathrm{i}}
\renewcommand{\Re}{\mathrm{Re}\,}
\renewcommand{\Im}{\mathrm{Im}\,}

\date{}
\author{}

\begin{document}

\title{The Gaussian free field and SLE$_4$ on doubly connected domains}
 \title{The Gaussian free field and SLE$_4$ on doubly connected domains}
 
 \maketitle

\begin{center}
\large Christian Hagendorf\footnote{Department of Physics, University of Virginia, 382 McCormick Road, Charlottesville, VA 22904-4714. {\small \tt
      hagendorf@virginia.edu}}, Denis Bernard\footnote{Member of the
    CNRS; Laboratoire de Physique Th\'eorique, Ecole Normale
    Sup\'erieure, 24 rue Lhomond, 75005 Paris, France.  {\small \tt
      denis.bernard@ens.fr}}, and 
	Michel Bauer\footnote{Institut de Physique
    Th\'eorique de Saclay, CEA Saclay, 91191 Gif-sur-Yvette, France
    and Laboratoire de Physique Th\'eorique, Ecole Normale
    Sup\'erieure, 24 rue Lhomond, 75005 Paris, France.  {\small \tt
      michel.bauer@cea.fr}}
\end{center}

\vspace{.3cm}
\begin{abstract}
	The level lines of the Gaussian free field are known to be related to SLE$_{4}$. It is shown how this relation allows to define chordal SLE$_4$ processes on doubly connected domains, describing traces that are anchored on one of the two boundary components. The precise nature of the processes depends on the conformally invariant boundary conditions imposed on the second boundary component. Extensions of Schramm's formula to doubly connected domains are given for the standard Dirichlet and Neumann conditions and a relation to first-exit problems for Brownian bridges is established. For the free field compactified at the self-dual radius, the extended symmetry leads to a class of conformally invariant boundary conditions parametrised by elements of SU$(2)$. It is shown how to extend SLE$_4$ to this setting. This allows for a derivation of new passage probabilities \`a la Schramm that interpolate continuously from Dirichlet to Neumann conditions.
\end{abstract}

\newpage

\section{Introduction}

The Schramm-Loewner evolutions (SLE) have proved to be a powerful tool to analyse the scaling limit of interfaces in two-dimensional systems of statistical mechanics at criticality. They constitute a one parameter family of conformally invariant planar growth processes that is obtained by solving Loewner's differential equation with a stochastic drift whose diffusion constant is a parameter $\kappa>0$ \cite{lawler:06}.

\begin{figure}[h]
  \centering
  \begin{tikzpicture}[scale=0.8]
 \filldraw[lightgray,xshift=-6cm] (0,0) circle (1.85cm);
  \draw[fill=white,xshift=-6cm] (0,0) circle (1.75cm);
  \filldraw[xshift=-6cm] (0,0) circle (\pointradius);

  \draw[xshift=-6cm] (0,0) node[right] {$0$};
  \draw[xshift=-6cm] (1.9,1.9) node {$\mathbb U$};
  \draw[xshift=-6cm] (1.9,0) node {$1$};
  \draw[xshift=-6cm]  (-1.5,1.05) node[left] {$e^{\i x}$};  

  \filldraw[color=black,xshift=-6cm] (1.75,0) circle (\pointradius);
  \filldraw[color=black,xshift=-6cm] (-1.475,0.95) circle (\pointradius);
  \draw[thick,dashed,yshift=-1cm,xshift=-6cm] (1.75,1) .. controls (1.45,1) and (1.5,0.6) .. (1.1,1) .. controls (0.7,1.5) and (1.3,1.8) .. (0.9, 2) .. controls (0,2.3) and (-0.4,2.5).. (-0.9,2.1) .. controls (-1.1,1.95) and (-1.1,2).. (-1.475,1.95);
  
  \draw[thick,xshift=-6cm] (1.75,0) .. controls (1,0) and (0.7,0) .. (0.3,-0.9) ..controls (0,-1.4) and (0,-1.8) .. (-1,-1.2) .. controls (-1.5,-0.8) and (-1.5,-0.2) .. (-0.9,0.4) .. controls (-0.5,0.7) and (-0.5,0.75) ..  (-1.475,0.95);
  \draw (-6,-2.5) node {(a)};
  \draw (0,-2.5) node {(b)};

  \filldraw[lightgray] (0,0) circle (1.85cm);
  \draw[thick,fill=white] (0,0) circle (1.75cm);
  \draw[thick,fill=lightgray] (0,0) circle (0.5cm);
  \filldraw[white] (0,0) circle (0.40cm);
 
  \draw (1.9,1.9) node {$\mathbb A_p$};
  \draw (1.9,0) node {$1$};

  \filldraw[color=black] (1.75,0) circle (\pointradius);
  \filldraw[color=black] (-1.475,0.95) circle (\pointradius);
  \draw[thick,dashed,yshift=-1cm] (1.75,1) .. controls (1.45,1) and (1.5,0.6) .. (1.1,1) .. controls (0.7,1.5) and (1.3,1.8) .. (0.9, 2) .. controls (0,2.3) and (-0.4,2.5).. (-0.9,2.1) .. controls (-1.1,1.95) and (-1.1,2).. (-1.475,1.95);
  
  \draw[thick] (1.75,0) .. controls (1,0) and (0.75,0) .. (0.3,-0.9) ..controls (0,-1.4) and (0,-1.8) .. (-1,-1.2) .. controls (-1.5,-0.8) and (-1.5,-0.2) .. (-0.9,0.4) .. controls (-0.5,0.7) and (-0.5,0.75) ..  (-1.475,0.95);
  
  \draw (-1.5,1.05) node[left] {$e^{\i x}$};
  \end{tikzpicture}
  \caption{(a) Left- and right-passage (solid and dashed line) for a trace from $1$ to $e^{\i x}$ with respect to $0$ on the unit disc, and (b) with respect to the inner boundary $|z|=e^{-p}$ on a circular annulus $\mathbb A_p$.}
  \label{fig:pprobs}
\end{figure}
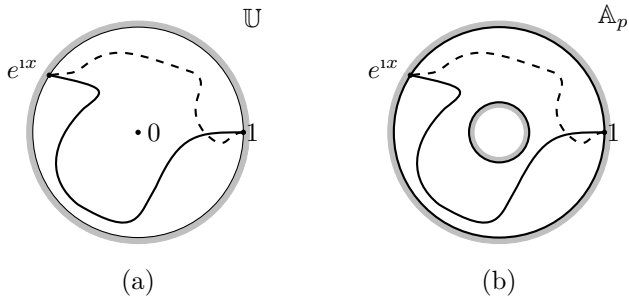

An application of the SLE theory is given by Schramm's formula: suppose that we consider interfaces on a simply connected domain joining two boundary points (chordal SLE$_\kappa$). Given some bulk point in the domain, we may ask for the probability that it lies to the left or to the right if we go along this interface from one boundary point to another. This problem was solved for any value of $\kappa$ \cite{schramm:01}. If we choose the disc geometry as depicted in figure \ref{fig:pprobs}(a), and chordal traces from $1$ to $e^{\i x}$ then the left-passage probability is given in terms of Gau{\ss}' hypergeometric function \cite{abramowitz:70}:
\begin{equation}
  \beta(x) =  \frac{1}{2}+\frac{\Gamma(4/\kappa)}{\sqrt{\pi}\,\Gamma((8-\kappa)/2\kappa)}\cot \left(\frac x 2\right)\,{}_2F_1\left(\frac{1}{2},\frac{4}{\kappa},\frac{3}{2};-\cot^2\left( \frac x 2\right)\right).
  \label{eqn:schramm}
\end{equation}

If we consider instead doubly connected domains, there is an obvious generalisation of Schramm's probability: a chordal SLE$_\kappa$ trace, whose ends are anchored on one boundary component, can circumvent the second boundary component in two ways as illustrated in figure \ref{fig:pprobs}(b) for a circular annulus $\mathbb A_p$ comprised between $|z|=e^{-p}$ and $|z|=1$. As a generalisation of \eqref{eqn:schramm} we would like to evaluate the left- or right-passage probability with respect to this boundary. As we may impose different (conformally invariant) boundary conditions on this boundary component the doubly connected case offers the possibility to study the influence of boundary conditions on the chordal SLE$_\kappa$ process. A simple example is provided by the Ising model: consider the annulus geometry of figure \ref{fig:pprobs}(b) and fix the Ising spins to $+$ on the boundary arc $(1,e^{\i x})$, and to $-$ on the complementary arc. Then there will be an interface joining $1$ and $e^{\i x}$ separating $+$ from $-$ spins whose scaling limit gives a SLE$_3$ process \cite{smirnov:09}. Intuitively, it is clear that the precise value of the left-passage probability with respect to the second boundary depends on our choice among the three conformally invariant boundary conditions imposed thereon: $+$, $-$ or free. Moreover, for free boundary conditions, we may even expect that the interface can hit this second boundary component with non-zero probability in the scaling limit.

As opposed to chordal SLE$_\kappa$ processes in simply connected domains the passage problem on doubly connected domains could so far not be solved for arbitrary values of $\kappa$.
Some explicit results were obtained for SLE$_2$, i.e. the scaling limit of the loop-erased random walk on doubly connected domains with Dirichlet boundary conditions \cite{hagendorf:09}, and Dirichlet-Neumann boundary conditions \cite{phdhagendorf}. The purpose of this article is to evaluate the left/right-passage and hitting probabilities for another solvable case: SLE$_4$ on doubly connected domains. The case $\kappa=4$ is known to be related to the level lines of the Gaussian free field which corresponds to a very simple conformal field theory (CFT) with central charge $c=1$. More precisely the SLE$_4$ trace is a discontinuity line emerging from finite discontinuities of the field that takes piecewise constant values at the boundary. The jump of the field across the discontinuity line takes a universal value, $2\lambda=\pi\sqrt{2}$ in our normalisations, irrespective of the precise value of the discontinuity at the boundary. Yet a proper SLE$_4$ trace is obtained only if the latter equals $2\lambda$, too. Otherwise, one has to deal with the so-called SLE$(4,\vec{\rho})$ processes \cite{cardy:04,schramm:09}. Hence, we concentrate on the free field theory on doubly connected domains like the circular annulus in figure \ref{fig:pprobs}(b), where the boundary values are $+\lambda$ on the arc $(1,e^{\i x})$ and $-\lambda$ on the complementary arc. Again, on the second boundary component we are free to choose any conformally invariant boundary condition.
We provide the passage probabilities for the canonical choices of Dirichlet and Neumann boundary conditions, and show that they are related to simple stochastic processes known as Brownian bridges and their generalisations. Moreover it turns out that our methods extend to the case of the compactified free field at the self-dual radius. The corresponding conformal field theory displays an extended symmetry, and is equivalent to the level $1$ Wess-Zumino-Witten model for SU$(2)$. We repeat our analysis for a large class of conformally-invariant boundary conditions parametrised by group elements of SU$(2)$, arising from marginal boundary deformations of the free field theory \cite{callan:94,callan:94_2,recknagel:99}. We thus derive passage probabilities that continuously interpolate from Dirichlet to Neumann boundary conditions while preserving conformal invariance.

The article is organised as follows. We start with a brief reminder about SLE on simply and doubly connected domains in section \ref{sec:slebasics}. In particular we explain the procedure of deforming SLE measures with martingales in order to intertwine between different versions of SLE. Our arguments rely on the correspondence between SLE martingales and ratios of CFT correlation functions \cite{bauer:02,bauer:03,bauer:05_2}. Therefore, in section \ref{sec:freefield} we recall elementary methods to compute boundary correlation functions for the Gaussian free field using Gaussian functional integrals as well as some standard regularisation schemes, and discuss their relation to SLE$_4$ martingales. We use these in order to
define chordal SLE$_4$ on doubly connected domains with different boundary conditions. In section \ref{sec:pprobs} we show how these SLE$_4$ processes allow to map the problem of left/right-passage on doubly connected domains or hitting of the second boundary component to a first-exit problem of Brownian bridges from an interval (here the open interval $(0,2\pi)$). For Dirichlet boundary conditions on the inner boundary the corresponding probabilities depend sensitively on the boundary values taken by the field, and a detailed analysis of the different cases is given in (\ref{eqn:resdirstart}-\ref{eqn:resdirend}). For Neumann boundary conditions the corresponding Brownian bridge lives on a circle, and we generalise the solution of the first-exit problem in (\ref{eqn:lpdn},\ref{eqn:hitdn}). Section \ref{sec:su2} treats the compactified free field at the self-dual radius. We recall some facts about generalised SU$(2)$ boundary conditions and then compute a (non-trivial) boundary two-point function on doubly connected domains for the boundary condition changing (b.c.c.) operators relevant to SLE. The two-point function allows to define some generalised SLE$_4$ variants. We find the passage probabilities and hitting probabilities for these SLE processes by solution of the corresponding first-exit problem. Our main results can be found in (\ref{eqn:su2hit1},\ref{eqn:su2hit2},\ref{eqn:su2lp}). Moreover, we discuss in detail the hitting probability and show that non-hitting of the inner boundary of the annulus is only possible for pure Dirichlet boundary conditions with special values taken by the field at the boundary. We present our conclusions in section \ref{sec:conclusion}. An appendix recalls some facts about SU$(2)$ current one-point functions on doubly connected domains and Lie derivatives.

\section{SLE on simply and doubly connected domains}
\label{sec:slebasics}
We start from some well-known facts about SLE on simply connected domains (for reviews see \cite{cardy:05,bauer:06}, or the book \cite{lawler:06}). Given a simply connected planar domain $\mathbb D$ and two boundary points $x_0,x_\infty\in \partial \mathbb D$ the \textit{chordal} Schramm-Loewner evolution is defined by a measure $\mathsf P_{\mathbb D,x_0,x_\infty}$ on non-self-crossing curves $\gamma$ from $x_0$ to $x_\infty$ in $\mathbb D$ with the following two properties. \textit{(i) Conformal transport:} if $f(z)$ is a conformal mapping from $\mathbb D$ to $\mathbb D' = f(\mathbb D)$, and $x'_0,x'_\infty$ the images $x_0,x_\infty$ then the image of a sample trace $\gamma$ has probability measure $\mathsf P_{\mathbb D',x'_0,x'_\infty}$; \textit{(ii) Domain Markov property:} suppose that we condition the curves on a first portion $\gamma_1$ from $x_0$ to $z\in \mathbb D$, then the remainder $\gamma\backslash \gamma_1$ from $z$ to $x_\infty$ has the distribution in the domain cut along $\gamma_1$, i.e. $\mathsf P_{\mathbb D,x_0,x_\infty}[\gamma|\gamma_1]= \mathsf P_{\mathbb D\backslash \gamma_1,z,x_\infty}[\gamma\backslash \gamma_1]$. If instead we are interested in curves joining a boundary point $x_0\in \partial \mathbb D$ and a bulk point $z\in \mathbb D$, we deal with a version called \textit{radial} SLE.

The built-in conformal invariance of SLE leaves us free to choose a reference domain of particular symmetry. As a reference domain for doubly connected domains we use a cylinder $\mathbb T_p$ that corresponds to the rectangle $\{z\in \mathbb C\mathop{|}0 \leq \Re z < 2\pi,\, 0 < \Im z < p\}$ with $z$ and $z+2\pi$ identified. The annulus $\mathbb A_p$ mentioned in the introduction is obtained through a conformal transformation $w = e^{\i z}$.
In fact, for every doubly connected domain $\mathbb D$ it exists a $p>0$ such that $\mathbb D$ may be conformally mapped onto $\mathbb T_p$ \cite{nehari:82}. The parameter $p$ is called \textit{modulus} and specifies a conformal equivalence class. We shall often call $\Im z = 0$ and $\Im z = p$ the lower and upper boundary of the cylinder.

\subsection{SLE on simply connected domains}

\paragraph{Loewner's equation, radial SLE.}  In the limit $p\to \infty$ we obtain a semi-infinite cylinder $\mathbb T_\infty$ which is a simply connected domain when completed with the point at infinity $\i\infty$. We consider traces $\gamma$ that start from $x_0=1$, and are parametrised by some time parameter $t\geq 0$. We denote by $\gamma_t$ the tip of the curve at time $t$, and by $\gamma_{[0,t]}=\bigcup_{0\leq s \leq t}\gamma_s$ the trace to time $t$. For simplicity, let us suppose for the moment that $\gamma_{[0,t]}$ has no double points (this condition has to be relaxed for general SLE processes, see below). Since $\mathbb T_\infty\backslash \gamma_{[0,t]}$ is still simply connected (at least for sufficiently small $t$) there is a conformal mapping $g_t(z)$ from this cut domain back to $\mathbb T_\infty$ thanks to the Riemann mapping theorem. The time evolution of this mapping is governed by \textit{Loewner's differential equation}. For $\mathbb T_\infty$ it is given by
\begin{equation}
	\frac{\diff g_t(z)}{\diff t} = \cot\left(\frac{g_t(z)-W_t}{2}\right),\quad g_{t=0}(z)=z.
	\label{eqn:loewnerradial}
\end{equation}
In fact, the time parametrisation was chosen so that the $g_t(z)\sim z + 2t/z+\dots$ as $\Im z\to +\infty$ (in fact, this is formally equivalent the standard normalisation for chordal SLE in the upper half plane \cite{lawler:06}). The real-valued function $W_t$ is the image of the tip $\gamma_t$ under $g_t(z)$ and encodes the full information about $\gamma_{[0,t]}$. For the radial case from $0$ to $\infty$ (here $\infty$ corresponds to the limit $\Im z\to +\infty$) conformal invariance and the domain Markov property (and reflection symmetry) lead to $W_t = \sqrt{\kappa}B_t$ where $B_t$ is standard Brownian motion, and $\kappa >0$ a positive real parameter. This corresponds to radial SLE (on an infinite cylinder).

Conversely, we may study traces resulting from injecting $W_t=\sqrt{\kappa}B_t$ into Loewner's equations \eqref{eqn:loewnerradial} what yields a conformally invariant growth process. For $0\leq \kappa \leq 4$ the traces $\gamma_{[0,t]}$ are simple curves; for $4< \kappa < 8$ they have double points and in this case $g_t(z)$ rather maps $\mathbb T_\infty\backslash \mathbb K_t$ back to $\mathbb T_\infty$ where $\mathbb K_t$ is the complement of the connected component of $\mathbb T_\infty\backslash \gamma_{[0,t]}$ that contains the point $\infty$, called the \textit{hull}. For $\kappa \geq 8$ the traces become space filling Peano curves \cite{rohde:05}. For several values of $\kappa$ the Schramm-Loewner evolution SLE$_\kappa$ describes interfaces in critical models of statistical mechanics: for example $\kappa=2$ corresponds to the scaling limit of the loop-erased random walk \cite{lawler:04}, $\kappa=8/3$ to the scaling limit of the self-avoiding walk (\cite{lawler:04bis}, still a conjecture), and $\kappa=8$ to the Peano curve of the uniform spanning tree \cite{lawler:04}.

\paragraph{Chordal SLE.} The chordal case \textit{on the infinite cylinder} $\mathbb T_\infty$ involves a slight complication. If we would like to study SLE$_\kappa$ traces from $0$ to $x$ then the presence of a marked point on the boundary implies that $W_t$ is not pure Brownian motion anymore. It rather has an additional drift \cite{schramm:05,sheffield:09}:
\begin{align}
  \label{eqn:driftsde}
	 \diff W_t &=\sqrt{\kappa}\,\diff B_t+ \left(\frac{6-\kappa}{2}\right)\cot\left(\frac{X_t-W_t}{2}\right)\,\diff t,\quad X_t=g_t(x).
\end{align}
This defines a chordal SLE process up to the first time where the two points $X_t$ and $W_t$ meet, what means that the growth process has reached its final point. This equation is obtained upon mapping the usual chordal SLE$_\kappa$ on the upper half-plane onto $\mathbb T_\infty$.  However, it is worth mentioning a more probabilistic approach using statistical mechanics martingales and the SLE-CFT-correspondence \cite{bauer:02,bauer:03}. To this end, consider the cut domain $\mathbb  T_{\infty;t} = \mathbb T_\infty\backslash \gamma_{[0,t]}$, and the partition functions $Z^{\mathrm{rad}}_{\mathbb T_{\infty;t}}(\gamma_t,\infty)$ and $Z^{\mathrm{chord}}_{\mathbb T_{\infty;t}}(\gamma_t,x)$ for radial traces from $\gamma_t$ to $\infty$, and chordal traces from $\gamma_t$ to $x$ respectively. In the scaling limit, we know that they correspond to CFT correlation functions $Z^{\mathrm{rad}}_{\mathbb U_t}(\gamma_t,\infty) = \langle \Phi_{0,1/2}|\psi_{1,2}(\gamma_t)\rangle_{\mathbb T_{\infty;t}}$ and $Z^{\mathrm{chord}}_{\mathbb T_{\infty;t}}(\gamma_t,x) = \langle \psi_{1,2}(\gamma_{t})\psi_{1,2}(x)\rangle_{\mathbb T_{\infty;t}}$. Here $\Phi_{r,s}$ and $\psi_{r,s}$ denote bulk and boundary primary fields with conformal weights $h_{r,s}=((\kappa r-4s)^2-(\kappa-4)^{2})/16\kappa$. In particular $\psi_{1,2}(x)$ is the b.c.c. operator whose insertion into correlation functions mimics the presence of an SLE trace anchored at the boundary point $x$, and $\Phi_{0,1/2}$ is the equivalent bulk operator \cite{bauer:04_3}. Using general arguments from statistical mechanics, one shows that \cite{bauer:02,bauer:03}
\begin{equation}
  M_{t} = \frac{Z^{\mathrm{chord}}_{\mathbb T_{\infty;t}}(\gamma_t,x)}{Z^{\mathrm{rad}}_{\mathbb T_{\infty;t}}(\gamma_t,\infty)} = \frac{|g_{t}'(x)|^{h_{1,2}}Z^{\mathrm{chord}}_{T_{\infty}}(W_{t},X_t)}{Z^{\mathrm{rad}}_{\mathbb T_\infty}(W_t,\infty)}
  \label{eqn:deformationmg}
\end{equation}
is a local martingale for radial SLE on $\mathbb T_\infty$, and corresponds to the Radon-Nykodim derivative of chordal SLE on $\mathbb T_\infty $ with respect to radial SLE \cite{bauer:09}. Here we have tacitly used the standard transformation formulae for primary fields under conformal mappings. Using this martingale, we can apply Girsanov's theorem \cite{oksendal:07} in order to deform the radial SLE measure to the chordal one, and work out the stochastic differential equation of chordal SLE on $\mathbb T_\infty$:
\begin{equation*}
  \diff W_{t} = \sqrt{\kappa}\, \diff B_{t} + \kappa \frac{\partial}{\partial w}\ln Z_{T_\infty}^{\mathrm{chord}}(w=W_{t},X_t)\diff t.
\end{equation*}
In order to obtain this equation we implicitly used that $Z_{\mathbb U}^{\mathrm{rad}}(W_t,\infty)$ is a constant due to translation invariance. The equivalence with \eqref{eqn:driftsde} then follows from the general form of boundary two-point functions $\langle \psi_{1,2}(x)\psi_{1,2}(y)\rangle_{\mathbb T_\infty} = \left|2\sin\left({(x-y)}/{2}\right) \right|^{-h_{1,2}}$, $h_{1,2}=(6-\kappa)/2\kappa$, which is fixed by global conformal invariance.

\subsection{SLE on doubly connected domains}
\label{sec:sledcd}
The generalisation of Loewner's equation to planar domains with arbitrary connectivity was first addressed by Komatu \cite{komatu:40,komatu:50}. Within the SLE context it has been investigated by Bauer and Friedrich \cite{rbauer:06,rbauer:08}, in particular the doubly-connected case was studied by Zhan \cite{zhan:04,zhan:06,zhan:09} (see also \cite{bauer:04_2} for a derivation based on symmetry arguments).

Loewner's equation on $\mathbb T_p$ is given by
\begin{equation}
	\label{eqn:loewnercyl}
	\frac{\diff g_t(z)}{\diff t} = v(g_t(z)-W_t,p-t),\quad  g_{t=0}(z)=z,
\end{equation}
where the vector field on the right-hand side is
\begin{equation}
	v(z,p) = \cot \left(\frac{z}{2}\right) + 4\sum_{n=1}^\infty\frac{\sin n z}{e^{2np}-1}.
	\label{eqn:defv}
\end{equation}
As $p\to +\infty$ while $t$ is kept finite we recover the radial Loewner equation \eqref{eqn:loewnerradial}.
The process $W_t$ is the image of the tip of the trace $\mathbb \gamma_t$. Since $\mathbb T_p\backslash \mathbb \gamma_{[0,t]}$ lies in a different conformal equivalence class than $\mathbb T_p$ the modulus changes in the course of the evolution. For the doubly connected case this evolution may be reabsorbed into the time parametrisation: in fact, the solutions to \eqref{eqn:loewnercyl} map $\mathbb T_p \backslash \mathbb \gamma_{[0,t]}$ onto $\mathbb T_{p-t}$. Therefore the evolutions ends for $t=p$ (at the latest). The growth process is determined by the time evolution of $W_t$. The standard choice $W_t = \sqrt{\kappa}B_t$, was analysed by Zhan \cite{zhan:04,zhan:06}: injection of this driving process into Loewner's equation yields a conformally invariant growth process. In this case for $\kappa\leq 4$ the traces emerge from $0$ and exit the cylinder at the upper boundary at $t=p$ (for $\kappa >4$ the tip of the trace may hit the lower boundary). For example $\kappa=2$ corresponds to the scaling limit of the loop-erased random walk on $\mathbb T_p$ from $0$ to the upper boundary. 
\begin{figure}
  \centering
  \begin{tikzpicture}[>=stealth]
  
     \filldraw[lightgray] (-1.5,2.5) rectangle (1.5,2.6);
	\filldraw[lightgray] (-1.5,0) rectangle (1.5,-0.1);
	\draw (-1.5,0) -- (1.5,0);
	\draw (-1.5,2.5) -- (1.5,2.5);
	\draw[dotted] (-1.5,0) -- (-1.5,2.5);
	\draw[dotted] (1.5,0) -- (1.5,2.5);
	\draw[<->] (-1.8,0) -- (-1.8,2.5);
	\draw (-2.1,1.25) node {$p$};
	
	\draw (-1.475,1.25) node[rotate=45] {$=$};
	\draw (1.525,1.25) node[rotate=45] {$=$};
	
	\draw (-1.5,-0.4) node {\small $0$};
	\draw (0.5,-0.4) node {\small $x$};
	\draw (1.5,-0.4) node {\small $2\pi$};
	
	\filldraw (-1.5,0) circle (\pointradius);
	\filldraw (0.5,0) circle (\pointradius);
	\filldraw (1.5,0) circle (\pointradius);
	
	\draw[densely dashed] (-1.5,0) .. controls (-1.5, 0.25) and (-1.,0.15) ..  (-0.95,0.5) .. controls (-0.9,0.75) and (-1.3,0.7) .. (-1.5,0.75);
	\draw[densely dashed] (1.5,0.75) .. controls (1.3,0.8) and (1.1,0.9) .. (1.1,1.2) .. controls (1.1, 1.5) and (1.1,1.7) .. (1.5,1.75);
	\draw[densely dashed] (-1.5, 1.75) .. controls (-0.75,1.9) and (-0.75,1.9) .. (0, 1.5) .. controls (0.5,1.2) and (0.75,1) .. (0.75,0.75) .. controls (0.75,0.5) and (0.5,0.5) .. (0.5,0);
	
	\draw[thick] (-1.5,0) .. controls (-1.5, 0.25) and (-1.,0.15) ..  (-0.95,0.5) .. controls (-0.95,0.65) and (-1,0.7).. (-1.3,0.725);
	\filldraw (-1.3,0.725) circle (\pointradius);
	\draw (-0.55,0.4) node {\small $\gamma_{[0,t]}$};
	\draw (-1.3,0.725) node[above] {\small $\gamma_t$};

	\draw (1,2.9) node {$\mathbb T_{p}$};
	
	\draw[->] (2,1) -- (4,1);
	\draw (3,1) node[above] {$g_t(z)$};

    \filldraw[lightgray] (4.5,2) rectangle (7.5,2.1);
	\filldraw[lightgray] (4.5,0) rectangle (7.5,-0.1);
	\draw (4.5,0) -- (7.5,0);
	\draw (4.5,2) -- (7.5,2);
	\draw[dotted] (4.5,0) -- (4.5,2);
	\draw[dotted] (7.5,0) -- (7.5,2);
	
	\draw (4.525,1) node[rotate=45] {$=$};
	\draw (7.525,1) node[rotate=45] {$=$};
	
	\draw (4.9,-0.4) node {\small $W_t$};
	\draw (6.2,-0.4) node {\small $X_t$};
	\draw (7.5,-0.4) node {\small $2\pi$};
	\draw[<->] (4.9,0.2) -- (6.1,0.2);
	\draw (5.5,0.4) node {\small $Y_t$};

	\filldraw (4.9,0) circle (\pointradius);
	\filldraw (6.1,0) circle (\pointradius);
	\filldraw (7.5,0) circle (\pointradius);
	
	\draw[<->] (7.8,0) -- (7.8,2);
	\draw (8,1) node[right] {$p-t$};
		
	\draw (7,2.4) node {$\mathbb T_{p-t}$};
	
	\draw[densely dashed] (4.9,0) .. controls (4.9,0.5) and (4.4,1.5) .. (4.8, 1.7) .. controls (5.2,1.9) and  (5.2,1.8) .. (5.7,1.55) .. controls (6.3,1.2) and (6.3,1.1) ..  (6.3,0.75) .. controls (6.3,0.25) and (6.1,0.5) ..  (6.1,0);

  \end{tikzpicture}
  \caption{Motion of points on lower boundary of the cylinder as induced by the Loewner flow. The tip $\gamma_t$ of the first portion of the trace $\gamma_{[0,t]}$ grown up to time $t$ is sent to $W_t$ whereas the point $x$ follows the flow according to $ X_t = g_t(x)$.}
  \label{fig:annulusflow}
\end{figure}
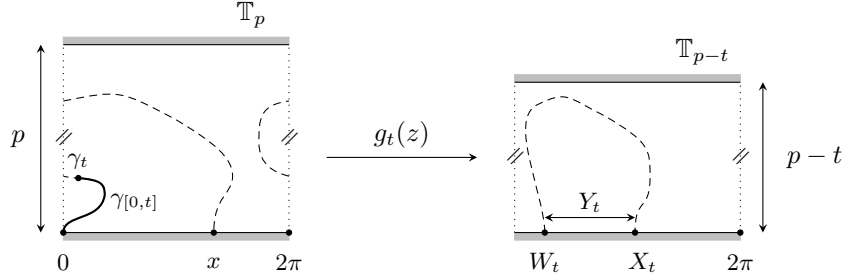

In the case of marked points like for chordal traces from $0$ to $x\in (0,2\pi)$ we have to add an appropriate drift to $W_t$ in order to force the growth process to terminate at $x$. Again, consider the cut domain $\mathbb T_{p;t} = \mathbb T_p\backslash \gamma_{[0,t]}$ and the partition functions $Z_{\mathbb T_{p;t}}(\gamma_t)$ and $Z^{\mathrm{chord}}_{\mathbb T_{p;t}}(\gamma_t,x)$ for the traces from $\gamma_t$ to the upper boundary and for chordal traces from $\gamma_t$ to $x$ respectively. The analogue of \eqref{eqn:deformationmg} is given by
\begin{equation}
  M_t = \frac{Z_{\mathbb T_{p;t}}^{\mathrm{chord}}(\gamma_t,x)}{Z_{\mathbb T_{p;t}}(\gamma_t)} = \frac{|g'_t(x)|^{h_{1,2}}Z_{\mathbb T_{p-t}}^{\mathrm{chord}}(W_t,X_t)}{Z_{\mathbb T_{p-t}}(W_t)},
  \label{eqn:cylmartingale}
\end{equation}
and has to be a martingale for the SLE variant with $W_t = \sqrt{\kappa}B_t$. Here we used the usual transformation rules for primary operators in conformal field theory, and the transformation properties of the conformal mapping $g_t(z)$. Because of translation invariance $Z_{\mathbb T_{p-t}}(W_t)$ cannot depend on $W_t$ and therefore is a function of $p-t$ alone; then a straightforward application of Girsanov's theorem yields the stochastic differential equation for chordal SLE on the cylinder:
\begin{equation*}
  \diff W_t = \sqrt{\kappa}\,\diff B_t	+ \kappa\frac{\partial}{\partial w}\ln Z_{\mathbb T_{p-t}}^{\mathrm{chord}}(w=W_{t},X_t)\,\diff t, \quad X_t = g_t( x).
\end{equation*}
In the sequel we consider problems that are invariant under global translations: the chordal partition function only depends on the relative coordinate $Y_t = X_t - W_t$ (see figure \ref{fig:annulusflow} for an illustration). It is convenient to abbreviate $Z_{\mathbb T_{p-t}}^{\mathrm{chord}}(W_t,X_t)= Z(Y_t,p-t)$. We then have to solve the stochastic differential equations
\begin{align}
	\diff W_t &= \sqrt{\kappa}\,\diff B_t-\kappa \frac{\partial}{\partial y}\ln Z(y=Y_t,p-t)\,\diff t, \label{eqn:slesde}
\\
	\diff Y_t &= -\diff W_t +v(Y_t,p-t)\diff t,
	\label{eqn:slesde2}
\end{align}
where $v(y,p)$ is the vector field from Loewner's equation. Actually, the second of these equations simply states that the end point $x$ follows the flow during the evolution and therefore is a direct consequence of \eqref{eqn:loewnercyl}. Upon insertion of \eqref{eqn:slesde} into \eqref{eqn:slesde2} we obtain a single stochastic differential equation for the stochastic process $Y_t$ which will be our starting point for the analysis of the passage problem (see below). As for the simply connected case, the chordal SLE process defined through these equations is defined only up the first time where $X_t$ and $W_t$ meet, i.e. when $Y_t = 0$ or $2\pi$, at the latest however at $t=p$.
We will see below that this simple statement allows to reformulate the passage properties of the traces as a first-exit problem for the process from the interval $(0,2\pi)$.
However, in order to fully characterise the chordal SLE process we must first compute the partition function $Z(y,p) = \langle \psi_{1,2}(y)\psi_{1,2}(0)\rangle_{\mathbb T_p}$. In contrast to the limit $p\to \infty$, there is no global conformal invariance fixing this correlation function. In the next section, we show how to find explicit expressions in the case of $\kappa=4$ from free field calculations.

\section{Free field boundary correlation functions}
\label{sec:freefield}
From now on we concentrate on the Gaussian free field/free boson on a domain $\mathbb D$. In two dimensions it corresponds to a conformal field theory with central charge $c=1$, defined through the action
\begin{equation*}
	S[X] = \frac{1}{2\pi}\int_{\mathbb D}\diff^2 z\, \partial X(z,\bar z)\bar \partial X(z,\bar z).
\end{equation*}
If the planar domain is finite, we must specify boundary conditions (that preserve the conformal symmetry). In this section we focus on two cases: \textit{(i)} for Dirichlet the field is (piecewise) constant $X(z,\bar z)\equiv\mathrm{const.}$ along the boundary; \textit{(ii)} for Neumann boundary conditions we have $\partial X(z,\bar z)/\partial \nu = 0$ at the boundary, where $\nu$ denotes the interior normal unit vector at $z\in \partial \mathbb  D$. Correlation functions can then be computed from the one- and two-point functions, and the Wick theorem. We shall refer to $X_{\mathrm{cl}}(z,\bar z)=\langle X(z,\bar z)\rangle_{\mathbb D}^{\mathrm{b.c.}}$ as the ``classical configuration'' which is the (essentially) unique harmonic function on $\mathbb D$, that obeys the prescribed boundary conditions.

\subsection{Simply connected domains}
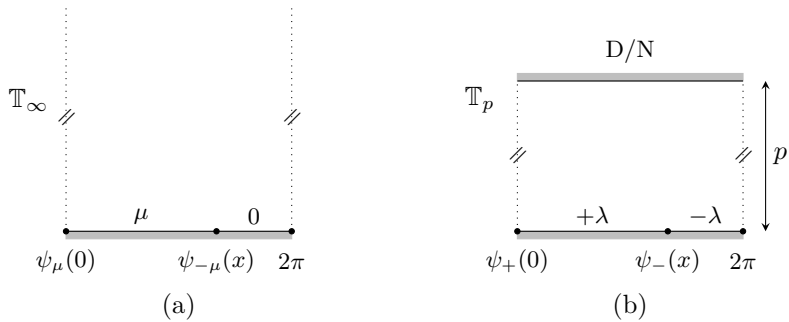
\begin{figure}[h]
  \centering
  \begin{tikzpicture}[>=stealth]
    \filldraw[white] (-1.5,-1.5) rectangle (9,3.5);
 
	\filldraw[lightgray] (-1.5,0) rectangle (1.5,-0.1);
	\draw (-1.5,0) -- (1.5,0);
	\draw[dotted] (-1.5,0) -- (-1.5,3);
	\draw[dotted] (1.5,0) -- (1.5,3);
	
	\draw (-1.475,1.5) node[rotate=45] {$=$};
	\draw (1.525,1.5) node[rotate=45] {$=$};
	\filldraw (-1.5,0) circle (\pointradius);
	\filldraw (0.5,0) circle (\pointradius);
	\filldraw (1.5,0) circle (\pointradius);
	
	\draw (-1.5,-0.4) node {\small $\psi_{\mu}(0)$};
	\draw (0.5,-0.4) node {\small $\psi_{-\mu}(x)$};
	\draw (1.5,-0.4) node {\small $2\pi$};
	
	\draw (-0.5,0.2) node {\small $\mu$};
	\draw (1,0.2) node {\small $0$};

	\draw (-2,1.75) node {$\mathbb T_\infty$};

	\filldraw[lightgray] (4.5,2) rectangle (7.5,2.1);
	\filldraw[lightgray] (4.5,0) rectangle (7.5,-0.1);
	\draw (4.5,0) -- (7.5,0);
	\draw (4.5,2) -- (7.5,2);
	\draw[dotted] (4.5,0) -- (4.5,2);
	\draw[dotted] (7.5,0) -- (7.5,2);
	
	\draw (4.525,1) node[rotate=45] {$=$};
	\draw (7.525,1) node[rotate=45] {$=$};
	
	\draw (4.5,-0.4) node {\small $\psi_+(0)$};
	\draw (6.5,-0.4) node {\small $\psi_-(x)$};
	\draw (7.5,-0.4) node {\small $2\pi$};
	
	\draw (5.5,0.2) node {\small $+\lambda$};
	\draw (7,0.2) node {\small $-\lambda$};
	
	\filldraw (4.5,0) circle (\pointradius);
	\filldraw (6.5,0) circle (\pointradius);
	\filldraw (7.5,0) circle (\pointradius);
	
	\draw[<->] (7.8,0) -- (7.8,2);
	\draw (8,1) node {$p$};
	
	\draw (6,2.4) node {\small D/N};
	
	\draw (4,1.75) node {$\mathbb T_p$};
		
		\draw (6,-1) node {(b)};
		\draw (0,-1) node {(a)};

  \end{tikzpicture}
  \caption{(a) Piecewise constant boundary conditions on the infinite cylinder. (b) Finite with discontinuous Dirichlet boundary conditions on the lower boundary. On the upper boundary we impose either Dirichlet or Neumann boundary conditions.}
  \label{fig:scdir}
\end{figure}

For a start it is instructive to consider the simply connected case $\mathbb T_\infty$ with piecewise constant Dirichlet boundary conditions on its boundary $\Im z = 0$ as shown in figure \ref{fig:scdir} for two discontinuities. Therefore consider $M$ points $0\leq x_1 < x_2 <\dots < x_M < 2\pi$ and suppose that the field has a discontinuity of $\mu_k$ when passing through $x_k$ along the boundary in positive direction. For the boundary conditions to be consistent we impose $\sum_{k=1}^M\mu_k=0$. The partition function $Z=Z(x_1,\dots,x_M)$ is the sum over all configurations subject to these boundary conditions weighted by $\exp - S[X]$, i.e. the functional integral
\begin{equation*}
  Z(x) = \int_{\mathrm{b.c.}}[\diff X]\, \exp - S[X].
\end{equation*}
The usual strategy for Gaussian functional integrals consists of dividing any configuration into the classical configuration and fluctuations $X(z,\bar z) = X_{\mathrm{cl}}(z,\bar z) + X_{\mathrm{fl}}(z,\bar z)$. From the fact that $X_{\mathrm{cl}}(z,\bar z)$ is harmonic and $X_{\mathrm{fl}}(z,\bar z)$ vanishes at the boundary we see that $S[X] = S[X_{\mathrm{cl}}]+S[X_{\mathrm{fl}}]$. The functional integral with respect to the fluctuations yields $1/\sqrt{\det -\Delta}$.

We write the classical configuration as $X_{\mathrm{cl}}(z,\bar z)=\Re f(z)$ where $f(z)$ has the ``mode expansion''
\begin{equation}
  f(z) = -\sum_{k=1}^M\frac{\mu_k x_k}{2\pi}-\sum_{k=1}^M\frac{\i \mu_k}{\pi}\sum_{n=1}^\infty \frac{e^{\i n (z-x_k)}}{n}.
  \label{eqn:classdisc}
\end{equation}
The classical action $S[X_{\mathrm{cl}}] = \int_{\mathbb T_\infty} \diff^2 z\,|f'(z)|^2 /8\pi$ is infinite because the sharp discontinuities at the boundary yield an infinite ``elastic energy''. From the point of view of mode expansions this divergence comes from ultraviolet modes with large $n$. We remedy this by truncation of the sum \eqref{eqn:classdisc} at $n=N$ and compute the action
\begin{align*}
 S_N[X_{\mathrm{cl}}]=H_N\sum_{k=1}^M\frac{\mu_k^2}{8\pi^2}-\sum_{k<l}\frac{\mu_k\mu_l}{4\pi^2}\,\sum_{n=1}^N\frac{\cos n (x_k-x_l)}{n}
\end{align*}
where $H_N$ is the $N$-th harmonic number, diverging according to $H_N\sim \ln N$ for large $N$ \cite{abramowitz:70}. Hence as $N\to \infty$ the partition function behaves to the leading order as
\begin{equation*}
  Z(x_1,\dots,x_M) = \frac{a^{\sum_{k=1}^M\mu_k^2/8\pi^2}}{\sqrt{\det -\Delta}}\prod_{k<l}\left|2\sin \left(\frac{x_l-x_k}{2}\right)\right|^{\mu_k\mu_l/4\pi^2}
\end{equation*}
where we introduced the short-distance cutoff $a = 1/N$. It is common lore to interpret $a^{-{\sum_{k=1}^M\mu_k^2/8\pi^2}}\sqrt{\det -\Delta}\,Z(x_1,\dots,x_M)$ as $M$-point function of primary b.c.c. operators in conformal field theory. Let us therefore introduce the operators $\psi_{\mu_k}(x)$ that shift the value of the field $X(z,\bar z)$ by $\mu_k$ if we pass through the boundary point $x$ in positive direction. Hence we find
\begin{equation*}
  \left\langle \prod_{k=1}^M\psi_{\mu_k}(x_k)\right\rangle_{\mathbb T_\infty} = \prod_{k<l}\left|2\sin \left(\frac{x_l-x_k}{2}\right)\right|^{\mu_k\mu_l/4\pi^2}.
\end{equation*}
From $M=2$ we have $\langle \psi_{-\mu}(0)\psi_{\mu}(x)\rangle_{\mathbb T_\infty}=|2\sin (x/2)|^{-\mu^2/4\pi^2}$ and thus read off that the operators $\psi_{\mu}$ have conformal weight $h = \mu^2/8\pi^2$. Moreover, from $M\geq 4$ we deduce fusion rules for these operators. Omitting the details we find
\begin{align}
  \psi_{\mu_1}\times \psi_{\mu_2}&\to \psi_{\mu_1+\mu_2}, \qquad \mu_1+\mu_2\neq 0 \label{eqn:fusion1},\\
  \psi_{-\mu}\times \psi_{\mu}&\to1+\frac{\partial X}{\partial \nu} \label{eqn:fusion2}.
\end{align}
Our normalisation is chosen so that the fusion coefficient in \eqref{eqn:fusion1} equals $1$. The same holds for fusion to the identity in \eqref{eqn:fusion2}. However, the fusion coefficient to the normal derivative of the field is given by $\mu/4\pi$.

For chordal SLE$_4$ on $\mathbb T_\infty$ the partition function is thus given by a two-point function of operators creating discontinuities of $\mu_\pm = \pm \pi\sqrt{2}$, as was shown in the mathematical literature \cite{schramm:09}. We abbreviate $\psi_\pm(x)= \psi_{\pm \pi\sqrt{2}}(x)$ and find the conformal weights $h_\pm = 1/4$ as well as the correlation function $\langle \psi_-(x)\psi_+(0)\rangle_{\mathbb T_\infty}= |2\sin (x/2)|^{-1/2}$ in accordance with the SLE-CFT correspondence.

\subsection{Doubly connected domains}
We now consider doubly-connected domains. We choose the cylinder $\mathbb T_p$ as reference geometry, and impose on its lower boundary $\Im z = 0$ Dirichlet conditions (see figure \ref{fig:scdir}(b)): the field takes the value $+\lambda$ on $(0,x)$ and $-\lambda$ on $(x,2\pi)$ with $\lambda=\pi/\sqrt{2}$. This amounts to study the insertion of $\psi_-(x)\psi_+(0)$ on a $-\lambda$-Dirichlet boundary into CFT correlation functions. For the upper boundary $\Im z = p$ we are free to choose any boundary conditions that preserves conformal invariance. Here we shall consider \textit{(i)} pure Dirichlet boundary conditions: $X\equiv \mu = \mathrm{const.}$ and \textit{(ii)} pure Neumann boundary conditions: $\partial X/\partial \nu=0$ where $\nu$ is the interior normal vector at some point of the upper boundary.
 
\subsubsection{Dirichlet-Dirichlet boundary conditions}
\label{sec:ddbc}
We write the classical configuration as $X_{\mathrm{cl}}(z,\bar z)=\Re f(z)$ with the complex-valued
function
\begin{equation*} 
	 f(z) =\frac{\lambda(x-\pi)}{\pi
}-\frac{\i\left(\lambda(x-\pi)-\pi \mu\right) z}{\pi p} +\frac{2\i\lambda}{\pi} \sum_{n\neq 0}
\frac{\left(e^{-\i n x}-1\right)e^{\i n z}}{n(1-e^{-2np})}. \label{eqn:dsol}
\end{equation*}
Following the same strategy as for the simply connected case we find 
to the partition function
\begin{equation}
	Z(x,p)= \frac{a^{\lambda^2/\pi^2}}{\sqrt{\det-\Delta}}\Bigg(\frac{\eta(\i p/\pi)^3\exp\left[-(x-\pi-\pi\mu/\lambda)^2/4p\right]}{
	\left|\theta_1\left({x/2\pi}|\i p/\pi\right)\right|}\Bigg)^{\lambda^2/\pi^2}
	\label{eqn:ddpf}
\end{equation}
where $\theta_1(z|\tau)=- \sum_{n\in \mathbb Z} e^{2\pi\i (n+1/2) (z+1/2)+\i \pi \tau (n+1/2)^2}$ is a Jacobi $\theta$-function, and Dedekind's $\eta$-function $\eta(\tau) = e^{i\pi \tau/12}\prod_{n=1}^\infty\left(1-e^{2\pi \i n\tau}\right)$.
The determinant of the Laplacian with Dirichlet-Dirichlet boundary conditions is given by $\left(\det -\Delta \right)_{\mathbb T_p}= p\,\eta(\i p/\pi)^2/\pi$ (this can be found from the well-known $\zeta$-function regularisation scheme or lattice calculations when properly removing the zero mode of the Laplacian). For the critical value $\lambda=\pi/\sqrt{2}$ \eqref{eqn:ddpf} coincides with the partition function we seek for. Hence we deduce the two-point boundary correlation function:
\begin{equation}
	\mathcal A\,\langle \psi_-(x)\psi_+(0)\rangle_{\mathbb T_p}= \left|\frac{\eta(\i p/\pi)}{
	\theta_1\left({x/2\pi}|\i p/\pi\right)}\right|^{1/2}\label{eqn:ddcf}
	\times \sqrt{\frac{\pi}{p}}\exp\left[-\frac{(x-\pi-\sqrt{2}\mu)^2}{8p}\right],
\end{equation}
where
\begin{equation}
	\mathcal A = \eta(\i p/\pi)^{-1}\,\sqrt{\frac{\pi}{p}}\exp\left[-\frac{(\pi+\sqrt{2}\mu)^2}{8p}\right]
	\label{eqn:ddam}
\end{equation}
is the CFT partition function (cylinder amplitude). Since the operators involved are b.c.c. operators, the normalisation of correlation functions is a bit subtle. Here we have normalised in such a way that as $x\to 0^+$ the correlation function behaves to leading order like $x^{-1/2}$ (with coefficient 1). In this limit the lower boundary of $\mathbb T_p$ corresponds to $-\lambda$-Dirichlet conditions as suggested by figure \ref{fig:scdir}(b). We will use this convention in the following sections.

\subsubsection{Dirichlet-Neumann boundary conditions}
\label{sec:dnbc}
We may alter the preceding
case by imposing Neumann boundary conditions on the upper boundary
$\Im z = p$. The classical configuration is given by
$X_{\mathrm{cl}}(z,\bar z)= \Re f(z)$ with the complex-valued function
\begin{equation*}
	f(z) = \frac{\lambda (x-\pi)}{\pi}+\frac{2\i \lambda}{\pi}\sum_{n\neq 0}^\infty\left(\frac{(e^{-\i n x}-1)e^{\i n z}}{n(1+e^{-2np})}\right).
\end{equation*}
We follow the same lines as above in order to compute the regularised partition function. A straightforward calculation gives
\begin{equation}
	Z(x,p) = \frac{a^{\lambda^2/\pi^2}}{\sqrt{\det -\Delta}}\left(\frac{\eta(2\i p/\pi)^2\theta_4(x/2\pi|2\i p/\pi)^2}{\eta(\i p/\pi)\left|\theta_1({x/2\pi}|\i p/\pi)\right|}\right)^{\lambda^{2}/\pi^2},
	\label{eqn:dnpf}
\end{equation}
where $\theta_4(z|\tau)=\sum_{n\in \mathbb Z} (-1)^n e^{2\pi \i n z+\i \pi \tau n^2}$ is yet another Jacobi $\theta$-function.
The determinant of the Laplacian with Dirichlet-Neumann boundary conditions is given by $\left(\det -\Delta\right)_{\mathbb T_p} = 2(\eta(2\i p/\pi)/\eta(\i p/\pi))^2$. Moreover, this result is equivalent to a boundary two-point function.
As for the Dirichlet-Dirichlet case we may write for the critical value $\lambda =  \pi/\sqrt{2}$ the boundary two-point function
\begin{align}
	\mathcal A\langle \psi_-(x)\psi_+(0)\rangle_{\mathbb T_p}
	&= \left|\frac{\eta(\i p/\pi)}{
	\theta_1\left({x/2\pi}|\i p/\pi\right)}\right|^{1/2}\frac{\theta_4(x/2\pi|2\i p/\pi)}{\sqrt 2}.
  \label{eqn:dncf}
\end{align}
with $\mathcal A$ being the CFT partition function for Dirichlet-Neumann boundary conditions
\begin{equation}
	\mathcal A = \frac{\theta_4(0|2\i p/\pi)}{\sqrt{2}\,\eta(\i p/\pi)}=\frac{1+2\sum_{n=1}^\infty (-1)^n e^{-n^2p}}{\sqrt{2}\,\eta(\i p/\pi)}.
	\label{eqn:dnam}
\end{equation}

\subsection{Null vector equations and SLE$_4$ martingales}
\label{sec:nve}

In order to check the coherence of our procedure, we must verify that insertion of the partition functions \eqref{eqn:ddpf} and \eqref{eqn:dnpf} in \eqref{eqn:cylmartingale} indeed yields a (local) martingale $M_t$.

\paragraph{Null vectors and two-point functions.} It is known that SLE martingales are related to CFT null vectors \cite{bauer:02,bauer:03}. The SLE-CFT correspondence predicts that the b.c.c. operator $\psi_{1,2}$ has a null vector $(\kappa L_{-1}^2 - 4L_{-2})\psi_{1,2}=0$ in its Verma module where $L_n$ are the modes of the stress tensor that obey the Virasoro algebra \cite{bauer:06,difrancesco:97}. This leads to differential equations for correlation functions involving this operator. In particular, for $\kappa = 4$ the cylinder boundary two-point function $\langle \psi_{1,2}(0)\psi_{1,2}(x)\rangle_{\mathbb T_p}$ must be solution of
\begin{align}
	\Biggl(2\frac{\partial^2}{\partial x^2} + v(x,p)&\frac{\partial}{\partial x}+\frac{1}{4}v'(x,p)\Biggr)\left(\frac{\mathcal A \langle \psi_{1,2}(0)\psi_{1,2}(x)\rangle_{\mathbb T_p}}{\eta(\i p/\pi)^{1/2}}\right)\nonumber\\
	&= \frac{\partial}{\partial p}\left(\frac{\mathcal A \langle \psi_{1,2}(0)\psi_{1,2}(x)\rangle_{\mathbb T_p}}{\eta(\i p/\pi)^{1/2}}\right)
	\label{eqn:level2pde}
\end{align}
In our specific case, we have the doublet $\psi_\pm$ which corresponds to two copies of the Virasoro representation associated to $\psi_{1,2}$. Hence the correlation functions \eqref{eqn:ddcf} and \eqref{eqn:dncf} should be solution of this partial differential equation what is readily checked by explicit calculation. Moreover, they hint at a particular structure: these two correlation functions are given as a product of $|\eta(\i p/\pi)/\theta_1(x/2\pi|\i p/\pi)|^{1/2}$ and a solution of the heat equation. Indeed one may show that the general ansatz
\begin{equation}
  \mathcal A \langle \psi_-(x)\psi_+(0)\rangle_{\mathbb T_p} = \left|\frac{\eta(\i p/\pi)}{\theta_1(x/2\pi|\i p/\pi)}\right|^{1/2}f(x,p)
  \label{eqn:gentwopoint}
\end{equation}
is solution to \eqref{eqn:level2pde} provided that $f(x,p)$ solves the simple heat equation ($\dot f$ indicates the derivative of $f$ with respect to $p$, the prime the derivative with respect to $x$)
\begin{equation}
  \dot f(x,p) = 2f''(x,p).
  \label{eqn:heat}
\end{equation}
Notice that besides the cylinder amplitude $\mathcal A$ it is the function $f(x,p)$ that encodes substantial information about both boundary conditions, and therefore cannot be determined from the sole Virasoro degeneracy of the b.c.c. operators living on the Dirichlet boundary.

\paragraph{An SLE$_4$ martingale.} Given these observations, we infer the structure of the martingales \eqref{eqn:cylmartingale} for Zhan's SLE$_\kappa$ on doubly connected domains in the case $\kappa=4$:
\begin{equation*}
  M_t = \frac{|g_t'(x)|^{1/4} f(Y_t, p-t)}{|\theta_1(Y_t/2\pi|\i(p-t)/\pi)|^{1/2}},\quad \dot f(x,p) = 2f''(x,p),
\end{equation*}
Recall that in this case $W_t = 2B_t$ and that the process $Y_t$ is solution of the stochastic differential equation
\begin{equation*}
  \diff Y_t = -2\diff B_t + v(Y_t,p-t)\diff t
\end{equation*}
with the function $v(y,p)$ defined in \eqref{eqn:defv}.
Checking that $M_t$ is a local martingale amounts to an application of It\^o's formula.

Using $M_t$ we can apply Girsanov's theorem according to the argument given in section \ref{sec:sledcd} and construct the driving process for the chordal SLE$_4$ processes that describe traces from $0$ to $x$. The process $W_t$ is then obtained from the stochastic differential equation \eqref{eqn:slesde}, and has in general some complicated form involving elliptic functions. Remarkably, these cancel out for the relative coordinate $Y_t$ which is solution of the equation
\begin{equation}
  \diff Y_t = -2\diff B_t + 4\,\frac{f'(Y_t,p-t)}{f(Y_t,p-t)}\diff t,\qquad  Y_{t=0} = x.
  \label{eqn:yprocess}
\end{equation}
Notice that because $f(x,p)$ solves the heat equation \eqref{eqn:heat} we may interpret $Y_t$ as a \textit{conditioned Brownian motion} (this is strongly reminiscent of a Doob $h$-transform of standard Brownian motion \cite{doob:83}, however with a function $h$ that carries an explicit time dependence). These observations will allow for a solution of the left-passage problem on doubly connected domains, and will provide a starting point for generalisation to the compactified free field.

\section{Passage probabilities for simply and doubly connected domains}
\label{sec:pprobs}
Having computed partition functions of the chordal SLE$_4$ traces on doubly connected domains we may now come back to our original problem and study the passage probabilities \`a la Schramm for the corresponding SLE processes. We proceed by first recalling the case of simply connected domains, in particular how the evaluation of these probabilities is related to an exit problem for a stochastic process from an interval. For $\kappa=4$ this process turns out to be just one-dimensional Brownian motion. This simplification, together with our analysis of the null vector equations and martingales on doubly connected domains for $\kappa=4$, suggests that a similar reduction to an elementary stochastic process might occur in the doubly connected geometry. Indeed, as we show in this section, we can recast the passage problem into a first-exit problem of simple Brownian bridges from an interval.

\subsection{Simply connected domains}
It is useful, and instructive to recall the basic strategy to find the passage probabilities in the limit $p\to +\infty$ of simply connected domains. For $\kappa=4$ we have $\langle\psi_-(x)\psi_+(0)\rangle_{\mathbb T_p}=|2\sin (x/2)|^{-1/2}$, and thus the driving process and the relative motion are solution to the stochastic differential equation
\begin{align*}
  \diff W_t = 2\,\diff B_t + \cot\left(\frac{Y_t}{2}\right)\diff t,\quad \diff Y_t = \cot\left(\frac{Y_t}{2}\right)\diff t - \diff W_t.
\end{align*}
Upon replacing the first into the second equation we see that $\diff Y_t = - 2\,\diff B_t$, what leads to $Y_t = x- 2\,B_t$. We conclude that for $\kappa=4$ the relative motion is just simple Brownian motion. Now, recall that the SLE process is defined only up to the time where $W_t$ and $g_t(x)$ meet, i.e. up to the stopping time $\tau = \inf \{t\mathop{:} Y_t \neq (0,2\pi)\}$. At $t=\tau$ the Loewner evolution has reached the final point $x$, and the conformal mapping uniformises the connected component of $\mathbb T_\infty\backslash \gamma_{[0,\tau]}$ to $\mathbb T_\infty$ while fixing the bulk point $\infty$. It follows that the trace goes to the left of $\infty$ if $Y_\tau = 2\pi$, and to the right if $Y_\tau = 0$. We define the right- and left-passage probabilities
\begin{align*}
  \alpha(x) = \mathsf P [Y_\tau = 0] \quad\text{and}\quad
  \beta(x) = \mathsf P [Y_\tau = 2\pi].
\end{align*}
These are subject to the sum rule $\alpha(x)+\beta(x) = 1$.
The solution to the exit problem for $Y_t$ from $(0,2\pi)$ is a standard exercise in stochastic processes \cite{oksendal:07}: notice that $M_t = Y_t = x - 2B_t$ is a martingale. On the one hand we have $\mathsf E[M_\tau] = M_0 = x$, on the other hand $\mathsf E[M_\tau] = 0\cdot \mathsf P[Y_\tau = 0] + 2\pi\cdot\mathsf P [Y_\tau = 2\pi] = 2\pi \beta(x)$. Thus, we find $\beta(x) = x/2\pi$, as expected from Schramm's formula \eqref{eqn:schramm} with $\kappa= 4$.

\subsection{Doubly connected domains}
The strategy to reformulate the computation of right-/left-passage probabilities for SLE processes, defined through \eqref{eqn:slesde}, on doubly connected domains as an exit problem for the relative motion $Y_t$ from $(0,2\pi)$ is straightforward. The process is defined up to the stopping time $\tau = \inf \{t: Y_t \notin (0,2\pi)\}\wedge p$ (the wedge in this notation indicates that the process stops at the latest at $t=p$ even if the process did not exit from the interval up to this time). If $\tau < p$ the growth process emanating from $0$ has reached the final point $x$, and the conformal mapping $g_\tau(z)$ uniformises the connected component of $\mathbb T_p\backslash \gamma_{[0,\tau]}$ that contains the upper boundary boundary to $\mathbb T_{p-\tau}$. It follows that the trace passes to the left of that boundary component if $Y_\tau = 2\pi$, and to the right if $Y_\tau = 0$. Hence as for the simply connected case we introduce the right- and left-passage probabilities
\begin{align}
  \alpha(x,p) = \mathsf{P}[\tau < p\;\mathrm{and}\;Y_\tau = 0], \text{ and }
  \beta(x,p) = \mathsf{P}[\tau < p\;\mathrm{and}\;Y_\tau = 2\pi].
    \label{eqn:lrprobs}
\end{align}
What happens if $\tau = p$? Since the Loewner evolution stops at this time we interpret $\tau=p$ as an event where the SLE trace hits the upper boundary of $\mathbb T_p$.
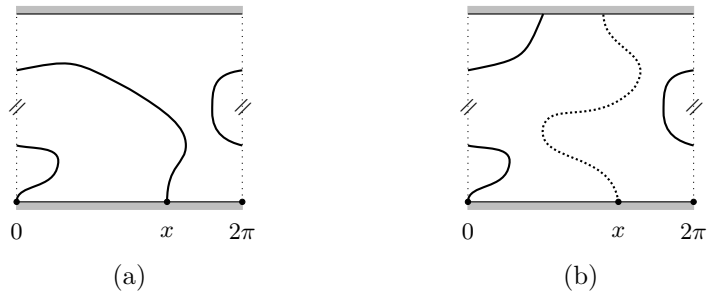
\begin{figure}
  \centering
  \begin{tikzpicture}[>=stealth]
  
    \filldraw[lightgray] (-1.5,2.5) rectangle (1.5,2.6);
	\filldraw[lightgray] (-1.5,0) rectangle (1.5,-0.1);
	\draw (-1.5,0) -- (1.5,0);
	\draw (-1.5,2.5) -- (1.5,2.5);
	\draw[dotted] (-1.5,0) -- (-1.5,2.5);
	\draw[dotted] (1.5,0) -- (1.5,2.5);
	
	\draw (-1.475,1.25) node[rotate=45] {$=$};
	\draw (1.525,1.25) node[rotate=45] {$=$};
	
	\draw (-1.5,-0.4) node {\small $0$};
	\draw (0.5,-0.4) node {\small $x$};
	\draw (1.5,-0.4) node {\small $2\pi$};
	
	\filldraw (-1.5,0) circle (\pointradius);
	\filldraw (0.5,0) circle (\pointradius);
	\filldraw (1.5,0) circle (\pointradius);
	
	\draw[thick] (-1.5,0) .. controls (-1.5, 0.25) and (-1.,0.15) ..  (-0.95,0.5) .. controls (-0.9,0.75) and (-1.3,0.7) .. (-1.5,0.75);
	\draw[thick] (1.5,0.75) .. controls (1.3,0.8) and (1.1,0.9) .. (1.1,1.2) .. controls (1.1, 1.5) and (1.1,1.7) .. (1.5,1.75);
	\draw[thick] (-1.5, 1.75) .. controls (-0.75,1.9) and (-0.75,1.9) .. (0, 1.5) .. controls (0.5,1.2) and (0.75,1) .. (0.75,0.75) .. controls (0.75,0.5) and (0.5,0.5) .. (0.5,0);	

\filldraw[lightgray] (4.5,2.5) rectangle (7.5,2.6);
	\filldraw[lightgray] (4.5,0) rectangle (7.5,-0.1);
	\draw (4.5,0) -- (7.5,0);
	\draw (4.5,2.5) -- (7.5,2.5);
	\draw[dotted] (4.5,0) -- (4.5,2.5);
	\draw[dotted] (7.5,0) -- (7.5,2.5);
	
	\draw (4.525,1.25) node[rotate=45] {$=$};
	\draw (7.525,1.25) node[rotate=45] {$=$};
	
	\draw (4.5,-0.4) node {\small $0$};
	\draw (6.5,-0.4) node {\small $x$};
	\draw (7.5,-0.4) node {\small $2\pi$};
	
	\filldraw (4.5,0) circle (\pointradius);
	\filldraw (6.5,0) circle (\pointradius);
	\filldraw (7.5,0) circle (\pointradius);

\draw[thick, xshift=6cm] (-1.5,0) .. controls (-1.5, 0.25) and (-1.,0.15) ..  (-0.95,0.5) .. controls (-0.9,0.75) and (-1.3,0.7) .. (-1.5,0.75);
\draw[thick,xshift=6cm] (1.5,0.75) .. controls (1.3,0.8) and (1.1,0.9) .. (1.1,1.2) .. controls (1.1, 1.5) and (1.1,1.7) .. (1.5,1.75);
\draw[thick,xshift=6cm] (-1.5, 1.75) .. controls (-0.75,1.9) and (-0.75,1.9) .. (-0.5, 2.5);
\draw [densely dotted,thick,xshift=6cm](0.5,0) .. controls (0.4,0.7) and (-0.6,0.5).. (-0.5,1) .. controls (-0.4,1.35) and (0.2,1) ..  (0.7,1.5) .. controls (1.05,2) and (0.3,2) .. (0.3,2.5); 	
   
   \draw (0,-1) node {(a)};
   \draw (6,-1) node {(b)};
  
  \end{tikzpicture}
  \caption{(a) Right-passage event of a trace on the cylinder corresponding to the event $Y_\tau =0$ with $\tau < p$. (b) If $\tau = p$ the trace hits the upper boundary. The same holds for a second trace anchored at $x$ which is not directly described by the SLE process.}
  \label{fig:events}
\end{figure}
Whether or not this event takes place depends sensitively on the boundary conditions that we impose on that boundary component. Hence we introduce the hitting probability
\begin{equation}
  \gamma(x,p) = \mathsf{P}[\tau=p].
  \label{eqn:hitprob}
\end{equation}
Because these three events exhaust all possible scenarios the probabilities must obey the sum rule
\begin{equation}
  \alpha(x,p) + \beta(x,p)+\gamma(x,p)=1.
  \label{eqn:sumrule}
\end{equation}
Figure \ref{fig:events} illustrates the two different cases of non-hitting vs. hitting schematically. From a microscopic point of view it becomes clear that in the case of hitting the upper boundary there is a second interface emanating from $x$ which also hits the upper boundary but is not directly described by the SLE process (this is very reminiscent of multiple SLE arch probabilities \cite{bauer:05}).

Having set this general framework we must solve the exit problem for the relative motion process $Y_t$. Our basic strategy consists of constructing martingales for the process $Y_t$ with suitable boundary conditions that project on the left-/right-passage or hitting events as $t=\tau$. Their expectation values then lead to the probabilities we are interested in.

\subsection{Dirichlet-Dirichlet boundary conditions}
Let us use the results from section \ref{sec:ddbc}. Combining \eqref{eqn:ddcf} and \eqref{eqn:yprocess} we find the stochastic differential equation:
\begin{equation}
	\diff Y_t = - 2\,\diff B_t+\left(\frac{Y_t-\pi-\sqrt{2}\mu}{p-t}\right)\diff t,\quad Y_{t=0}=x.
	\label{eqn:bbridge}
\end{equation}
This is the stochastic differential equation for a \textit{Brownian bridge} from $x$ to $y=\pi+\sqrt{2}\mu$, i.e. Brownian motion starting from $x$ and conditioned to visit $y$ at time $t=p$ \cite{oksendal:07}. Thus the passage problem for the SLE$_4$ traces with Dirichlet-Dirichlet boundary conditions on doubly connected domains is reduced to a first-exit problem for a one-dimensional Brownian bridge from the interval $(0,2\pi)$.

The exit probabilities \eqref{eqn:lrprobs} and \eqref{eqn:hitprob} depend sensitively on the final position $y$. The first assertion we can make is if $y \notin [0,2\pi]$ then the Brownian bridge leaves the interval with probability $1$, because $\lim_{t\to p^-}Y_t=y=\pi+\sqrt{2}\mu$ with probability $1$ \cite{oksendal:07}, at some $\tau < p$. Therefore, in this case $\gamma \equiv 0$. To proceed further we construct two martingales for the process $Y_t$ with appropriate boundary conditions which project at the different events as $t=\tau$ in a suitable manner. Since $Y_t$ is a conditioned Brownian motion we may write
	\begin{equation*}
	  M^{(k)}_t = \frac{h_k(Y_t,p-t)}{e^{-{(Y_t-y)^2}/{8(p-t)}}/\sqrt{8\pi(p-t)}},\quad k=1,2,
	\end{equation*}
	where $h_k(x,p)$ solves the heat equation $\dot{h}_k(x,p) = 2 h''_k(x,p)$.

\noindent

We use $M_{t}^{(1)}$ to construct the left-passage probability $\beta(x,p)$. Since $\beta(x,p)$ takes values $0$ and $1$ at $x=0$ and $x=2\pi$ respectively we shall impose the boundary conditions $h_1(0,p)=0$ and $h_1(2\pi,p) =e^{-{(2\pi-y)^2}/{8p}}/\sqrt{8\pi p}$ for all $p>0$. We can easily construct an arbitrary solution that fulfils the \textit{first} boundary condition by using the method of images. For $y\geq 0$ a reasonable guess, motivated from the well-known method of images, might be:
	\begin{equation*}
	  h_1(x,p) \mathop{=}^{?} \frac{1}{\sqrt{8\pi p}}\left(e^{-{(x-y)^2}/{8p}}-e^{-{(x+y)^2}/{8p}}\right).
	\end{equation*}
	The problem with this function is that it does not match the \textit{second} boundary condition at $x=2\pi$ because the second term leads to an unwanted term. We eliminate it by adding $e^{-{(x-y-4\pi)^2}/{8p}}/\sqrt{8\pi p}$. However, this spoils the boundary condition at $x=0$ what in turn can be readjusted upon subtracting $e^{-{(x+y+4\pi)^2}/{8p}}/\sqrt{8\pi p}$. Iteration of this procedure leads to
\begin{equation*}
	  h_1(x,p) = \frac{1}{\sqrt{8\pi p}}\sum_{n=0}^\infty \left(e^{-{(x-y-4\pi n)^2}/{8p}}-e^{-{(x+y+4\pi n)^2}/{8p}}\right),
\end{equation*}
and one can easily check that it obeys the desired boundary conditions. Moreover, the series is easily seen to be convergent for any $x\in (0,2\pi)$ and $y\geq 0$. A similar strategy can be applied to find a second martingale $M^{(2)}_t$ which we use to derive $\alpha(x,p)$: we seek for a solution $h_2(x,p)$ of the heat equation which takes values $h_2(0,p)=e^{-{y^2}/{8p}}/\sqrt{8\pi p}$ and $h_2(2\pi,p)=0$ for all $p>0$. Using similar methods as before we find for $y\leq 2\pi$:
\begin{equation*}
	  h_2(x,p) = \frac{1}{\sqrt{8\pi p}}\left(e^{-{(x-y)^2}/{8p}}{+}\sum_{n=1}^\infty \left(e^{-{(x-y+4\pi n)^2}/{8p}}{-}e^{-{(x+y-4\pi n)^2}/{8p}}\right)\right).
\end{equation*}
Using these two functions, it is not difficult to show that the local martingales $M_t^{(1)}$ and $M_t^{(2)}$ are bounded for $x\in [0,2\pi]$ and $y\geq 0$ and $y\leq 2\pi$ respectively for $t\leq \tau$. Therefore, they are proper martingales. At the stopping time they are given as sums of projectors
\begin{align*}
  M_{\tau}^{(1)} &= \bm 1_{\{\tau<p\;\mathrm{and}\;Y_\tau=2\pi\}}+(1-\delta_{y,0})\bm 1_{\tau = p},\quad y\geq 0,\\
  M_{\tau}^{(2)} &= \bm 1_{\{\tau<p\;\mathrm{and}\;Y_\tau=0\}}+(1-\delta_{y,2\pi})\bm 1_{\tau = p}\quad y\leq 2\pi,
\end{align*}
where we have used the Kronecker symbol $\delta_{x,y}=1$ for $x=y$, and $0$ otherwise, and the indicator function $\bm 1_A$ for some event $A$. Taking expectations (with respect to the Brownian bridge measure) we find
\begin{align}
  \beta(x,p) +&(1-\delta_{y,0})\gamma(x,p)\nonumber\\&= e^{{(x-y)^2}/{8p}}\sum_{n=0}^\infty \left(e^{-{(x-y-4\pi n)^2}/{8p}}-e^{-{(x+y+4\pi n)^2}/{8p}}\right)
  \label{eqn:firstpeqn}
\end{align}
for $y\geq 0$, and conversely for $y\leq 2\pi$
\begin{align}
  \alpha(x,p) +&(1-\delta_{y,2\pi})\gamma(x,p)\nonumber\\&= 1+e^{{(x-y)^2}/{8p}}\sum_{n=1}^\infty \left(e^{-{(x-y+4\pi n)^2}/{8p}}-e^{-{(x+y-4\pi n)^2}/{8p}}\right).
  \label{eqn:secondpeqn}
\end{align}
As $y\notin [0,2\pi]$ these formulae provide directly the left- and right-passage probabilities for $y> 2\pi$ and $y < 0$, respectively, because $\gamma\equiv 0$ as stated above. The missing probabilities can be obtained from the sum rule $\alpha(x,p)+\beta(x,p)=1$ in either case.

Conversely, for $y\in [0,2\pi]$ both equations are valid. Using the sum rule $\alpha(x,p)+\beta(x,p)+\gamma(x,p)=1$ we find
\begin{align}
  (1-\delta_{y,0}&-\delta_{y,2\pi})\gamma(x,p)\nonumber \\
  & = e^{{(x-y)^2}/{8p}}\sum_{n=-\infty}^\infty\left(e^{-{(x-y-4\pi n)^2}/{8p}}-e^{-{(x+y+4\pi n)^2}/{8p}}\right).\label{eqn:gamma}
\end{align}
This determines $\gamma(x,p)$ unless $y=0,2\pi$. Therefore, if $y\neq 0,2\pi$ we may insert it in \eqref{eqn:firstpeqn} and \eqref{eqn:secondpeqn}, and obtain $\alpha(x,p)$ and $\beta(x,p)$. If however $y=0$ or $2\pi$ then \eqref{eqn:gamma} does not define $\gamma$ (one verifies that -- consistently -- the right-hand side vanishes in these cases). Yet, since for times $t$ very close to $p$ the process oscillates around $y$ we conclude that even in this case $\gamma\equiv 0$.

\medskip

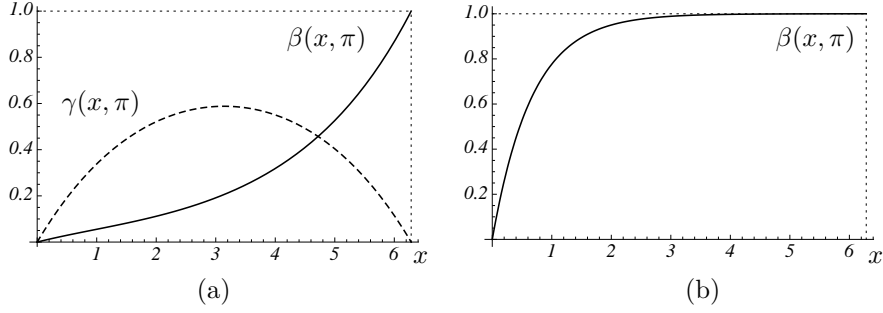
\begin{figure}[t]
	\centering
	\begin{tikzpicture}
	\draw (6.05,0) node {\includegraphics[width=0.45\textwidth]{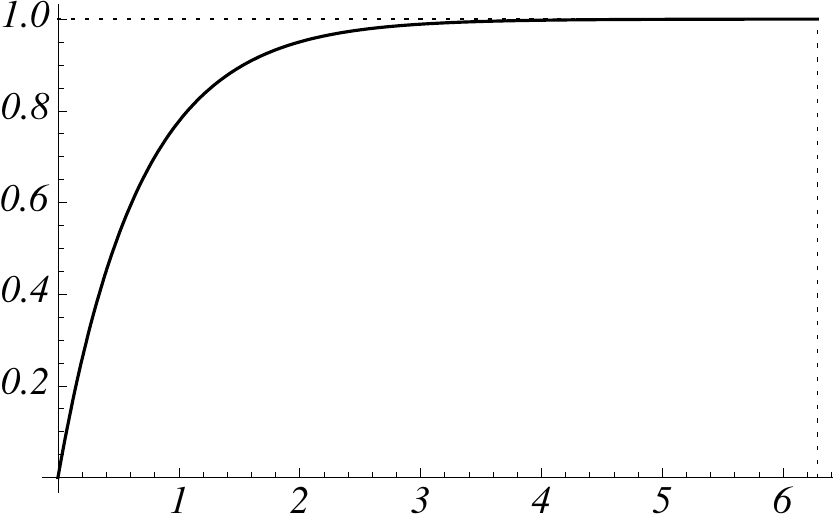}};
	\draw (0,0) node {\includegraphics[width=0.45\textwidth]{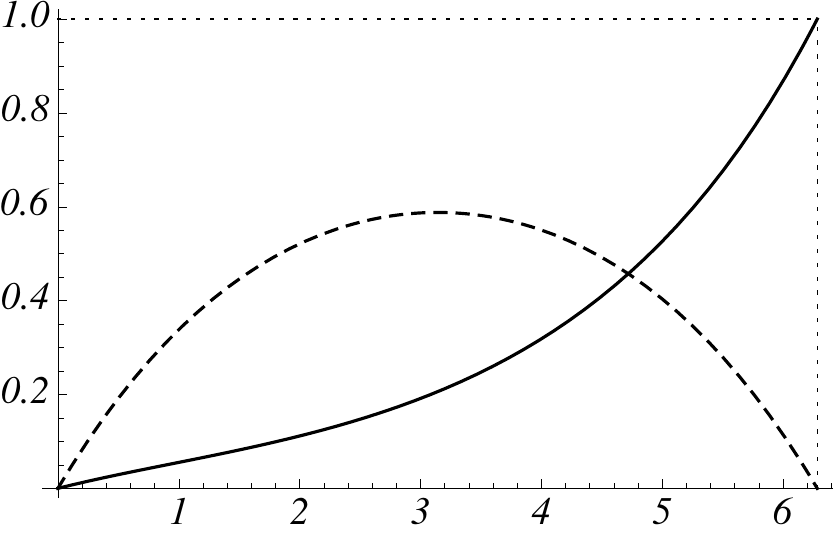}};
	\draw (2.75,-1.65) node {$x$};
	\draw (8.0,1.25) node {$\beta(x,\pi)$};
	\draw (1.5,1.25) node {$\beta(x,\pi)$};	
	\draw (-1.5,0.4) node {$\gamma(x,\pi)$};
	\draw (8.8,-1.65) node {$x$};
	
	\draw (0,-2.1) node {(a)};
	\draw (6.5,-2.1) node {(b)};

    \end{tikzpicture}
\caption{Illustration of the passage probabilities $\beta(x,p)$ and $\gamma(x,p)$ for Dirichlet-Dirchlet boundary conditions for the modulus $p=\pi$. (a) $\mu=0$ ($y=\pi(1+\mu/\lambda)=\pi$), (b) $\mu = 2\lambda$ ($y=\pi(1+\mu/\lambda)=3\pi$).}
\label{fig:ddprobs}
\end{figure}

\noindent
Let us now gather our results and list them explicitly for the different cases:
\begin{enumerate}
  \item$\bm{\,y\leq 0:}$ In this regime, the value of the field $\mu$ at the upper boundary is smaller that $-\lambda$. The hitting probability $\gamma$ vanishes identically. The passage probabilities are given by
  \begin{align}
   \alpha(x,p)& =1{-}e^{(x-y)^2/8p}\sum_{n=1}^\infty\left(e^{-{(x+y-4\pi n)^2}/{8p}}-e^{-{(x-y+4\pi n)^2}/{8p}}\right),\label{eqn:resdirstart}\\
    \beta(x,p) &= e^{{(x-y)^2}/{8p}}\sum_{n=1}^\infty\left(e^{-{(x+y-4\pi n)^2}/{8p}}-e^{-{(x-y+4\pi n)^2}/{8p}}\right).
\end{align}
  \item $\bm{y\in (0,2\pi):}$ If the value $\mu$ is comprised between $-\lambda$ and $+\lambda$ then the hitting probability $\gamma$ is non-zero. We find:
  \begin{align}
   \alpha(x,p)& =1{-}e^{(x-y)^2/8p}\sum_{n=0}^\infty\left(e^{-{(x-y-4\pi n)^2}/{8p}}-e^{-{(x+y+4\pi n)^2}/{8p}}\right),\\
    \beta(x,p) &= e^{{(x-y)^2}/{8p}}\sum_{n=1}^\infty\left(e^{-{(x+y-4\pi n)^2}/{8p}}-e^{-{(x-y+4\pi n)^2}/{8p}}\right),\\
    \gamma(x,p) &= e^{{(x-y)^2}/{8p}}\sum_{n=-\infty}^\infty\left(e^{-{(x-y-4\pi n)^2}/{8p}}-e^{-{(x+y+4\pi n)^2}/{8p}}\right)
.
\end{align}
  An illustration of the probabilities $\beta$ and $\gamma$ for $y=0$, i.e. for a Dirichlet value $\mu =0$, is given in figure \ref{fig:ddprobs}. 
  
  \item $\bm{y\geq 2\pi:}$ Here $\mu$ exceeds $+\lambda$. Again, the hitting probability $\gamma$ vanishes identically. The passage probabilities are given by
  \begin{align}
   \alpha(x,p)& =1{-}e^{(x-y)^2/8p}\sum_{n=0}^\infty\left(e^{-{(x-y-4\pi n)^2}/{8p}}-e^{-{(x+y+4\pi n)^2}/{8p}}\right),\\
    \beta(x,p) &= e^{{(x-y)^2}/{8p}}\sum_{n=0}^\infty\left(e^{-{(x-y-4\pi n)^2}/{8p}}-e^{-{(x+y+4\pi n)^2}/{8p}}\right)\label{eqn:resdirend}.
   \end{align}
   An illustration of the left-passage probability $\beta$ for $y=3\pi$ ($\mu =2\lambda$), is given in figure \ref{fig:ddprobs}.
\end{enumerate}
Finally we give an intuitive argument for the absence of hitting as $|\mu|\geq \lambda$: the SLE process that we have constructed describes the discontinuity line which is created from the jumps of the values taken by the field on the boundary. Loosely speaking if we cross this line, the field varies from $-\lambda$ to $+\lambda$ in a microscopic neighbourhood. Thus, as long as $-\lambda <\mu <\lambda$ the trace can get microscopically close to the upper boundary of the cylinder whereas this is not possible for values of $\mu$ outside this interval.

\subsection{Dirichlet-Neumann boundary conditions}
\label{sec:ndcase}
The Dirichlet-Neumann case may be worked out in the same way as before, now using the results from section \ref{sec:dnbc}.
Upon explicit evaluation of \eqref{eqn:yprocess} with the help of \eqref{eqn:dncf} we find
\begin{equation}
	\diff Y_t =  - 2\diff B_t+\frac{2\,\theta_4'(Y_t/2\pi|2\i (p-t)/\pi)}{\pi\,\theta_4(Y_t/2\pi|2\i (p-t)/\pi)}\,\diff t,\quad Y_{t=0}=x.
	\label{eqn:angledn}
\end{equation}
What type of stochastic process does it correspond to? To answer this question, consider Brownian motion $B_{4t}$ with $B_{t=0}=x$ and ask for the probability (density) that it is found at $\pi(2n+1)$ with some $n\in \mathbb Z$ at time $t=p$. We find $\mathsf P_x[B_{4p}\in [\pi(2n+1),\pi(2n+1)+\diff y] \text{ for some }n\in \mathbb Z] = \theta_4(x/2\pi|2\i p /\pi)\diff y$. It follows that \eqref{eqn:angledn} describes Brownian motion starting from $x$ and conditioned to visit $\pi(2n+1)$ for some $n\in \mathbb Z$ at time $t=p$. This is equivalent to a Brownian bridge on $S^1$, with $n$ playing the role of the winding number.

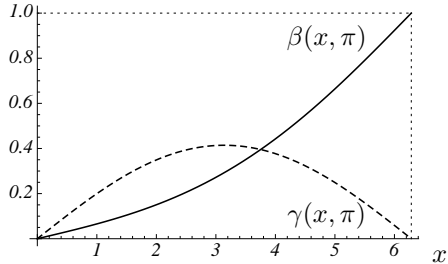
\begin{figure}
	\centering
	\begin{tikzpicture}
		\draw (0,0) node{\includegraphics[width=0.45\textwidth]{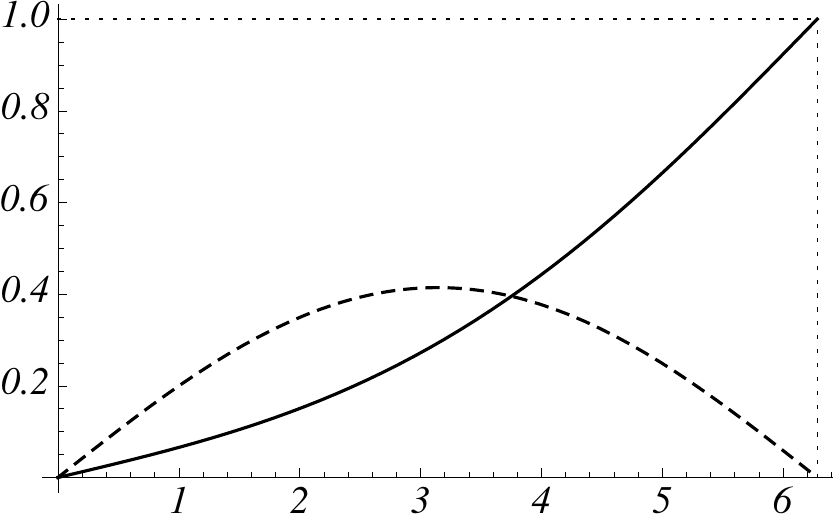}};
		\draw (3,-1.65) node {$x$};
		\draw (1.5,1.25) node {$\beta(x,\pi)$};	
		\draw (1.5,-1.1) node {$\gamma(x,\pi)$};
	\end{tikzpicture}
	\caption{Dirichlet-Neumann boundary conditions: the left-passage probability $\beta(x,p)$ (solid line), and the hitting probability $\gamma(x,p)$ (dashed line), both for $p=\pi$.}
	\label{fig:dnprobs}
\end{figure}

In order to determine the passage probabilities \eqref{eqn:lrprobs} and \eqref{eqn:hitprob} for the SLE$_4$ trace, let us consider the two functions
\begin{align*}
	u_1(y,p)&= \frac{\theta_1(y/2\pi,2\i p/\pi)}{\theta_4(y/2\pi,2\i p/\pi)}\\
	u_2(y,p)&= \frac{1}{2\pi}\left(\frac{2p\,\theta_4'(y/2\pi,2\i p/\pi)}{\pi\,\theta_4(y/{2\pi},2\i p/\pi)}+y\right).
\end{align*}
Using It\^o's formula one readily checks that $M^{(1)}_t=u_1(Y_t,p-t)$ and $M^{(2)}_t=u_2(Y_t,p-t)$ are local martingales for the process $Y_t$ defined through \eqref{eqn:angledn}. The idea behind the second martingale is to use the infinitesimal Lie symmetry $v=4p\,\partial_y+y\,u\partial_u$ of the heat equation $\dot u =2 u''$ \cite{olver:93}. For all $p>0$ we have $u_1(0,p) = u_1(2\pi,p) = 0,\, u_2(0,p) = 0$, and $u_2(2\pi,p)=1$. Moreover for $y\in (0,2\pi)$ we have the limits $\lim_{p\to 0^+}u_1(y,p)= 1$ and $\lim_{p\to 0^+}u_2(y,p)= 1/2$.
Using these boundary conditions, we can rewrite the martingales at the stopping time $t=\tau$ in terms of projectors $M_{\tau}^{(1)}=\bm{1}_{Y_\tau \in (0,2\pi)}$ and $M_{\tau}^{(2)}=\bm{1}_{Y_\tau = 2\pi}+\bm{1}_{Y_\tau \in (0,2\pi)}/2$. Hence, using the strong Markov property we find $\beta(x,p)= \mathsf E\left[M^{(1)}_\tau\right] = u_1(x,p)$ and $\gamma(x,p) = \mathsf E\left[M^{(2)}_\tau-M^{(1)}_\tau/2\right]= u_2(x,p)-u_1(x,p)/2$, or more explicitly
\begin{align}
	\beta(x,p) &= \frac{\sum_{n\in \mathbb Z}n\left(e^{-{(x-\pi (4n+1))^2}/{8p}}+e^{-{(x-\pi (4n-1))^2}/{8p}}\right)}{\sum_{n\in \mathbb Z} \left(e^{ -{(x-\pi(4n+1))^2}/{8p}}+e^{ -{(x-\pi (4n-1))^2}/{8p}}\right)\label{eqn:lpdn}
},\\
	\gamma(x,p) &= \frac{\sum_{n\in \mathbb Z}\left(e^{-{(x-\pi(4n +1))^2}/{8p}}-e^{-{(x-\pi(4n-1))^2}/{8p}}\right)}{\sum_{n\in \mathbb Z} \left(\,e^{ -{(x-\pi (4n+1))^2}/{8p}}+e^{ -{(x-\pi (4n-1))^2}/{8p}}\right)\label{eqn:hitdn}
},
\end{align}
These are the hitting and the left-passage probability for Dirichlet-Neumann boundary conditions. The two probabilities are illustrated on figure \ref{fig:dnprobs}. As before, the right-passage probability $\alpha(x,p)$ follows from the sum rule.

\section{Compactified free field and SLE$_4$ variants}
\label{sec:su2}

Up to now we have analysed the case of an uncompactified free field whose relation to SLE is well established \cite{schramm:09}. Yet the key observation in section \ref{sec:nve} is only based on \textit{(i)} $\kappa=4,c=1$ and \textit{(ii)} the fact, that the model contains a boundary operator with a Virasoro degeneracy at level two of its Verma module. This remains true if we compactify the free field at its self-dual radius $R=\sqrt{2}$ (in our normalisation). In this section we concentrate on this self-dual point where the theory is equivalent to the SU$(2)$ Wess-Zumino-Witten model at level $1$, i.e. one of the most simple conformal field theories with extended symmetry. For these CFTs it remains an open question to identify the geometric nature of the SLE traces \cite{bettelheim:05}. In the present case however, the connection with the Gaussian free field leads to the natural assumption that we may still think of them as discontinuity lines of the field, despite the compactification. We shall make this our working hypothesis, however this point deserves further investigation.

\subsection{Generalities, SU$(2)$ boundary states}
Upon compactification $X\equiv X+2\pi R$ the conformal field theory contains primary operators $\varphi_{n,m}(z,\bar z)$ with conformal weights
\begin{equation*}
  h = \frac{1}{2}\left(\frac{n}{R}+\frac{m R}{2}\right)^2, \quad \bar h = \frac{1}{2}\left(\frac{n}{R}-\frac{m R}{2}\right)^2,\quad n,m\in \mathbb Z,
\end{equation*}
and the self-dual radius is given by $R=\sqrt{2}$. At any rational multiple of the self-dual radius $R= \sqrt{2}M/N$, where $M,N$ are coprime integers, the theory contains a chiral Virasoro representation with conformal weight $h=1/4$ as can be seen from Bezout's lemma \cite{difrancesco:97}. Moreover, at $R=\sqrt{2}$ there are three chiral fields with conformal weight $1$: if we decompose the free boson into left- and right-movers $X(z,\bar z) = \phi(z)+\bar\phi(\bar z)$ then the currents are given by ($J^\pm(z)=J^1(z)\pm \i J^2(z)$)
\begin{equation}
	J^3(z) = \frac{\i\partial \phi(z)}{\sqrt{2}},\quad J^{\pm}(z) = :\exp \pm\i\sqrt{2}\phi(z):,
	\label{eqn:defcurrents}
\end{equation}
and likewise for the second copy (here $:\, :$ denotes the usual normal ordering). The currents form two copies of the $\widehat{\mathfrak{su}}(2)_1$ Kac-Moody algebra. The enhanced symmetry at this point makes the fields organise in spin-$j$ representations of SU$(2)$ with $j=0,1/2,1,3/2,\dots$ In particular, the boundary fields $\psi_\pm$ with conformal weights $h=1/4$ belong to the spin-$1/2$ representation (hence the fundamental representation). Let us point out that the spin-$1/2$ boundary operator was used by Bettelheim \textit{et al.} \cite{bettelheim:05} to extend the SLE approach to conformal field theories with $\widehat{\mathfrak{su}}(2)_k$ symmetries.

The currents $J^a(z)$ can be used to construct a large class of conformally invariant boundary conditions parametrised by $g\in \mathrm{SU}(2)$. In radial quantisation these are represented by boundary states $|g\rangle$ which are solution to the gluing conditions \cite{gaberdiel:01,gaberdiel:02}
\begin{equation}
	\left(\ad_{g\cdot \iota}(J^a_n)+\bar{J}_{-n}^a\right)|g\rangle = 0,\qquad
	\iota =
	\left(
	\begin{array}{cc}
		0 & 1\\
		-1 & 0
	\end{array}
	\right),
	\label{eqn:genbs}
\end{equation}
where $J^a_n,\bar J^a_n$ are the modes of the $\mathrm{SU}(2)$-currents, and $\ad_g (J^a_n) = gJ^a_n g^{-1}= \sum_b J^b_n\,(\ad\, g)^{ba}$. Given the structure of the gluing conditions we may interpret $|g\rangle$ as an $\mathrm{SU}(2)$-rotated Cardy identity state
\begin{equation}
	|g\rangle = (g\cdot \iota)|1\rangle_{\mathrm{Cardy}} .
	\label{eqn:rotatecardy}
\end{equation}
From the invariance of the Killing form we see that $|g\rangle$ respects the reparametrisation invariance $(L_n-\bar L_{-n})|g\rangle=0$ and therefore are proper conformally invariant boundary conditions. Moreover, the boundary states defined through \eqref{eqn:genbs} are known to be related to marginal boundary deformations of the free boson conformal field theory: these correspond to adding to the action periodic boundary interactions proportional to $:\sin X/\sqrt{2}:$ and $:\cos X/\sqrt{2}:$ with coupling constants that allow to determine $g$ and the boundary states $|g\rangle$ \cite{callan:94,callan:94_2,recknagel:99}. As an example, let us explicitly write the $\mathrm{U}(1)$-boundary states. The choice
\begin{equation*}
  g = \left(
  \begin{array}{cc}
    e^{\i \mu/\sqrt{2}} & 0\\
    0 & e^{-\i \mu/\sqrt{2}} 
  \end{array}
  \right)
\end{equation*}
leads to Dirichlet boundary conditions with $X\equiv \mu$ at the boundary. Conversely, Neumann boundary conditions are obtained with
\begin{equation*}
  g = \left(
  \begin{array}{cc}
    0 & e^{\i \tilde\mu/\sqrt{2}} \\
    -e^{-\i \tilde\mu/\sqrt{2}} & 0
  \end{array}
  \right),
\end{equation*}
where $\tilde \mu$ parametrises the so-called Wilson line (value taken by the dual field at the boundary). 

\subsection{Partition functions, current one-point functions}
\label{sec:pfc1pt}
Our goal consists of computing the two-point boundary function $\langle \psi_-(x)\psi_+(0)\rangle_{\mathbb T_p}$ for general SU$(2)$ boundary conditions. To this end, we need some additional input related to the cylinder amplitude and the one-point functions of the SU$(2)$ currents which we gather in this section.

The cylinder amplitude $\mathcal A = \mathcal A(g_1,g_2)$, which is the CFT partition function for the model with a $g_1$ boundary condition on the lower and a $g_2$ boundary condition on the upper boundary of the cylinder, is determined in terms of an angle $\alpha$ that is found from
\begin{equation*}
  2 \cos \alpha = \tr \left(g_1^{-1}g_2\right).
\end{equation*}
Here the trace is taken in the fundamental representation of SU$(2)$.
Since $\alpha$ is only determined modulo $2\pi$ we may restrict it to $\alpha\in [0,2\pi)$. Moreover, for any such $\alpha$ we have $2\pi-\alpha$ as further solution. However, in this section this ambiguity is not important since the expressions involving $\alpha$ are the same for either choice. In terms of this angle the cylinder amplitude is given by \cite{gaberdiel:01}
\begin{equation}
  \mathcal A = \langle g_1|e^{-p(L_0+\bar L_0 -1/12)}|g_2\rangle = \frac{1}{\sqrt{2}\,\eta(\i p/\pi)}\sum_{n\in \mathbb Z} e^{- n^2 p/2}\cos n\alpha.
  \label{eqn:partfunc}
\end{equation}
As an example, $-\lambda$-Dirichlet conditions on the lower boundary and $\mu$-Dirichlet conditions on the upper boundary conditions lead to $\alpha=\pi/2 + \mu/\sqrt{2}$. This can be seen from the uncompactified cylinder amplitude \eqref{eqn:ddam} after periodisation in $\mu$ with period $2\pi R = 2\sqrt{2}\pi$ (this corresponds to a summation over all winding sectors). The Dirichlet-Neumann case yields $\alpha=\pi/2$ what is compatible with \eqref{eqn:dnam} (the Dirichlet-Neumann partition functions for $R=\infty$ and $R=\sqrt{2}$ coincide \cite{callan:94_2}). More generally we impose $-\lambda$-Dirichlet boundary conditions on the lower boundary, and an arbitrary boundary condition on the upper boundary. The corresponding matrices are given by
\begin{equation}
  g_1 = \left(
    \begin{array}{cc}
      -\i & 0\\
      0 & \i
    \end{array}
  \right),\quad g_2 = \left(
    \begin{array}{cc}
      a & b\\
      - b^\ast & a^\ast
    \end{array}
  \right),\quad \mathrm{with} \quad |a|^2 + |b|^2 =1,
  \label{eqn:gmatrices}
\end{equation}
and the angle $\alpha$ is solution to $\cos \alpha = -\Im a$. As we may choose an arbitrary $g_2\in \mathrm{SU}(2)$, the parameter space of our two-boundary problem is given by the sphere $S^3$.

Besides the partition function we shall need the one-point functions of the current $J^3(z)$ on the cylinder. The general one-point function $\langle J^a(z)\rangle_{\mathbb T_p}$ is related to Lie derivatives of the partition function $\mathcal A$: $ \mathcal A(g_1,g_2)\langle J^a(z) \rangle_{\mathbb T_p} = \mathcal L^a_1 \mathcal A(g_1,g_2)$ along $t^a =\sigma^a/2$ where $\sigma^a$ denote the Pauli matrices (the subscript $1$ emphasises that the derivative acts on $g_1$). As we show in appendix \ref{app:lie} for $a=3$ this leads to
\begin{equation}
  \mathcal A \langle J^3(z)\rangle_{\mathbb T_p} = \frac{\Re a}{2\sqrt{2}\,\eta(\i p/\pi)\,\sin \alpha}\sum_{n\in \mathbb Z} n \,e^{n^2p/2}\,\sin n\alpha.
  \label{eqn:j3onept}
\end{equation}

\subsection{A boundary two-point function on the cylinder}
\label{sec:bdcfsu2}
We now compute the boundary two-point function $\langle \psi_-(x) \psi_+(0)\rangle_{\mathbb T_p}$ for SU$(2)$ boundary conditions characterised by the matrices given in \eqref{eqn:gmatrices}. Our analysis is based on a simple fusion argument: as $x\to 0^+$ we use the fusion rule \eqref{eqn:fusion2} and thus find
\begin{equation}
  \mathcal A \langle \psi_-(x) \psi_+(0)\rangle_{\mathbb T_p}\sim \mathcal A\, x^{-1/2} + \frac{ \mathcal A}{2\sqrt{2}}\left\langle \frac{\partial X(0)}{\partial \nu}\right\rangle_{\mathbb T_p} x^{1/2}+\dots
  \label{eqn:opetwopt}
\end{equation}
Since the normal derivative of the field $X$ is evaluated on a Dirichlet boundary, we can convert it to a SU$(2)$ current: $\langle{\partial X(0)}/{\partial \nu}\rangle_{\mathbb T_p}=2\sqrt{2}\langle J^3(0)\rangle_{\mathbb T_p}$, and evaluate the one-point function with the help of \eqref{eqn:j3onept}.
On the other hand, the general form \eqref{eqn:gentwopoint} suggests that
\begin{equation}
  \mathcal A \langle \psi_-(x) \psi_+(0)\rangle_{\mathbb T_p} \sim\frac{ f(0,p)}{\eta(\i p/\pi)} x^{-1/2} + \frac{f'(0,p)}{\eta(\i p/\pi)} x^{1/2}+\dots
  \label{eqn:opegentwopt}
\end{equation}
Here we used that $f(x,p)$ cannot be singular as $x\to 0^+$ in order to be consistent with the operator product expansion, and furthermore the series expansion $\theta_1(x/2\pi|\i p/\pi) \sim \eta(\i p/\pi)^3x+\dots$ for small $x$. Comparison of \eqref{eqn:opetwopt} and \eqref{eqn:opegentwopt} then leads to the boundary conditions
\begin{equation}
  f(0,p) = \eta(\i p/\pi)\mathcal A, \qquad f'(0,p) = \eta(\i p/\pi)\mathcal A\langle J^3(0)\rangle_{\mathbb T_p}. \label{eqn:boundaryconditions}
\end{equation}
We know that the function $f(x,p)$ has to be a solution of the heat equation $\dot f(x,p) = 2f''(x,p)$ as a consequence of the level two Virasoro degeneracy of the boundary operators $\psi_\pm$. The general expressions in \eqref{eqn:partfunc} and \eqref{eqn:j3onept} thus suggest that we should write this function as a Fourier series expansion
\begin{equation*}
  f(x,p) = \frac{a_0}{2} + \sum_{n=1}^\infty e^{-n^2 p/2}\left(a_n \cos\left( \frac{n x}{2}\right) + b_n \sin \left(\frac{n x}{2}\right)\right).
\end{equation*}
We determine the Fourier coefficients from \eqref{eqn:boundaryconditions}: the result is $a_n=\sqrt{2}\cos n \alpha$ and $b_n = \sqrt{2}\Re a\, \sin n\alpha/\sin \alpha$. More explicitly the function $f(x,p)$ takes the form
\begin{align}
   f(x,p)=\frac{1}{\sqrt{2}} +\sqrt 2\sum_{n=1}^\infty e^{-n^2p/2}\Biggl(&\cos n\alpha \cos \left(\frac{n x}{2}\right)+\frac{\Re a\,\sin n\alpha }{\sin\alpha }\sin \left(\frac{n x}{2}\right)\Biggr).
	\label{eqn:ffunction}
\end{align}
Hence we have succeeded to determine the correlation function $\langle \psi_-(x)\psi_+(0)\rangle_{\mathbb T_p}$ \eqref{eqn:gentwopoint} for generalised SU$(2)$ boundary conditions at the upper boundary of the cylinder.

\subsection{SLE$_4$ variants}
\subsubsection{Definition}
We use our result \eqref{eqn:ffunction} for the compactified free field at the self-dual radius in order to \textit{define} some SLE$_4$ variants from the partition function/two-point function $\langle \psi_-(x)\psi_+(0)\rangle_{\mathbb T_p}$ according to \eqref{eqn:slesde}.
As in the uncompactified case we study the process through the relative coordinate $Y_t$, whose time evolution is given by \eqref{eqn:yprocess}
\begin{equation*}
  \diff Y_t = -2\diff B_t + \frac{4f'(Y_t,p-t)}{f(Y_t,p-t)}\,\diff t,\quad Y_0=x.
\end{equation*}
We mentioned before that the structure of this stochastic differential equation implies that $Y_t$ is some conditioned (or Doob $h$-transformed) Brownian motion. What kind of conditioning? To give meaning to \eqref{eqn:yprocess} in this particular case it is useful to apply the Poisson summation formula and rewrite the function $f(x,p)$, given in \eqref{eqn:ffunction}, as
\begin{align*}
  f(x,p) =& \sqrt{\frac{\pi}{p}}\sum_{n\in \mathbb Z} \left(\omega_+\,e^{ -{(x-2\alpha-4\pi n)^2}/{8p}}+ \omega_-\,e^{ -{(x+2\alpha-4\pi n)^2}/{8p}}\right)
\end{align*}
with
\begin{equation}
  \omega_\pm= \frac{\sin \alpha\pm\Re a}{2\sin \alpha}, \quad \omega_++\omega_- = 1.
  \label{eqn:defomega}
\end{equation}
It is not difficult to show that for any SU$(2)$ boundary condition $g_2$ on the upper boundary we have $0\leq \omega_\pm \leq 1$. Therefore $f(x,p)$ is strictly positive for all $p>0$. If we introduce the two lattices $\mathcal L_\pm = \{\pm 2\alpha+4\pi n\}_{n\in \mathbb N}$ then we may interpret our process $Y_t$ as Brownian motion starting from $Y_0=x$, and conditioned to visit the lattice $\mathcal L_\pm$ at time $t=p$ with respective probability $\omega_\pm$. We thus encounter a generalisation of the process for Dirichlet-Neumann boundary conditions studied in section \ref{sec:ndcase}.

\subsubsection{Passage probabilities}
\begin{figure}[h]
  \centering
  \begin{tikzpicture}
    \draw (0,0) -- (2,0);
    \draw[|-|,very thick] (2,0) -- (5,0);
    \draw[-|] (5,0) -- (8,0);
    \draw (8,0) -- (10,0);
    \draw (2,-0.1) node[below] {$0$};
    \draw (5,-0.1) node[below] {$2\pi$};
    \draw (8,-0.1) node[below] {$4\pi$};
    \filldraw (2.75,0) circle (1.5pt);
    \filldraw[fill=white] (1.25,0) circle (1.5pt);
    \filldraw[fill=white] (7.25,0) circle (1.5pt);
    \filldraw (8.75,0) circle (1.5pt);
    \draw (1.25,0.33) node {\small $-2\alpha$};
    \draw (2.75,0.33) node {\small $2\alpha$};
     \draw (7.25,0.33) node {\small $4\pi-2\alpha$};
   \draw (8.9,0.33) node {\small $4\pi+2\alpha$};
   \draw (4,-0.1) -- (4,0.1);
   \draw (4,0.3) node {$x$};
   \draw [dotted] (10,0)-- (10.5,0);
   \draw [dotted] (-0.5,0)-- (0,0);
    
  \end{tikzpicture}
  \caption{Illustration for the diffusion process $Y_t$. It starts from $Y_0=x \in (0,2\pi)$ and visits the lattices $\mathcal L_\pm$, indicated by $\bullet$ and $\circ$ respectively, with probability $\omega_\pm$. In order to compute the passage/hitting probabilities we study the first-exit of this process from $(0,2\pi)$.}
  \label{fig:illdiff}
\end{figure}
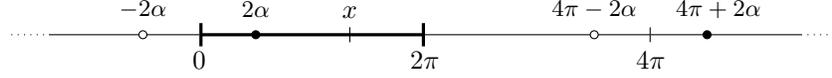

In order to find the passage probabilities \eqref{eqn:lrprobs} and \eqref{eqn:hitprob} for the SLE$_4$ traces on the cylinder, we must solve the first-exit problem from the interval $[0,2\pi]$ for the process $Y_t$ (see figure \ref{fig:illdiff} for an illustration).
The time evolution of the relative coordinate depends on the two complex parameters $a,b$ of the SU$(2)$ boundary condition on the upper boundary only through $\alpha$ and $\omega_\pm$. These are fully specified by the complex parameter $a$. The only restriction on $b$ then is given by the condition $|a|^2+|b|^2=1$. This equation determines only the modulus of the complex number $b$ but not its phase. It turns out that for given $a$ any of the probabilities $\alpha(x,p),\,\beta(x,p)$ and $\gamma(x,p)$ will be invariant under the replacement $b\to e^{\i \theta}b$.

As for the uncompactified free field we find the probabilities by constructing (local) martingales with appropriate boundary conditions and check their behaviour at the stopping time $\tau = \inf\{t\mathop{:} Y_t \notin (0,2\pi)\}\wedge p$. Since $Y_t$ is a conditioned Brownian motion, any such martingale will be of the form
\begin{equation*}
  M_t = \frac{h(Y_t,p-t)}{f(Y_t,p-t)}, 
\end{equation*}
with $h(x,p)$ being some solution of the heat equation $\dot h(x,p) = 2h''(x,p)$. Let us introduce two particular solutions which will be useful to solve our problem. The strategy outlined here below is a straightforward generalisation of the Dirichlet-Neumann case we have encountered in section \ref{sec:ndcase}.

\paragraph{Hitting probability.} Let us first consider $0\leq \alpha\leq \pi$. The function
\begin{equation*}
  h_1(x,p) = \omega_+ \sqrt{\frac{\pi}{p}}\sum_{n\in \mathbb Z}\left(e^{-{(x-2\alpha-4\pi n)^2}/{8p}}-e^{-{(x+2\alpha-4\pi n)^2}/{8p}}\right)
\end{equation*}
clearly is a solution of the heat equation.
It is positive for $x\in (0,2\pi)$ and for $p>0$ we have $h(x=0,p)=h(x=2\pi,p)=0$. Hence the function
\begin{equation*}
  u_1(x,p) = \frac{h_1(x,p)}{f(x,p)}
\end{equation*}
is positive for $x\in (0,2\pi)$ and zero at $x=0,2\pi$. Moreover, it is easy to show that for $x\in (0,2\pi)$ we have $\lim_{p\to 0^+} u_1(x,p) = 1$, and $u_1(0,p)=u_1(2\pi,p)=0$ for all $p>0$. Therefore $M^{(1)}_t = u_1(Y_t, p-t)$ is a local martingale for the process defined through \eqref{eqn:yprocess}. It is not difficult to check that it is bounded for $0\leq t \leq \tau$ and therefore a proper martingale. Moreover at the stopping time we have
\begin{equation*}
  M^{(1)}_\tau  = \bm 1_{Y_\tau \neq 0,2\pi}.
\end{equation*}
Therefore we find the probability that the trace hits the inner boundary $\gamma(x,p) = \mathsf E[M_\tau^{(1)}] = u_1(x,p)$, for $\alpha \in [0,\pi]$, or more explicitly
\begin{equation}
\gamma(x,p) = \frac{\omega_+ \sum_{n\in \mathbb Z}\left(e^{-{(x-2\alpha-4\pi n)^2}/{8p}}-e^{-{(x+2\alpha-4\pi n)^2}/{8p}}\right)}{\sum_{n\in \mathbb Z} \left(\omega_+\,e^{ -{(x-2\alpha-4\pi n)^2}/{8p}}+ \omega_-\,e^{ -{(x+2\alpha-4\pi n)^2}/{8p}}\right)
\label{eqn:su2hit1}
}.
\end{equation}
For $\pi \leq \alpha \leq 2\pi$ the analysis is similar and yields
\begin{equation}
\gamma(x,p) = \frac{\omega_- \sum_{n\in \mathbb Z}\left(e^{-{(x+2\alpha-4\pi n)^2}/{8p}}-e^{-{(x-2\alpha-4\pi n)^2}/{8p}}\right)}{\sum_{n\in \mathbb Z} \left(\omega_+\,e^{ -{(x-2\alpha-4\pi n)^2}/{8p}}+ \omega_-\,e^{ -{(x+2\alpha-4\pi n)^2}/{8p}}\right)\label{eqn:su2hit2}
}.
\end{equation}

In section \ref{sec:pprobs} we observed that for SLE$_4$ processes defined from the uncompactified free field the trace almost surely does not hit the upper boundary of the cylinder for Dirichlet boundary conditions with value $\mu$ if $|\mu|\geq \lambda$. Let us analyse this in the present situation. Non-hitting is equivalent to $\gamma\equiv 0$. First, for $\alpha \in [0,\pi]$ we must have $\omega_+=0$ what amounts to $\Re a = -\sin \alpha$ as can be seen from \eqref{eqn:defomega}. However, recall from section \ref{sec:pfc1pt} that $\Im a=-\cos \alpha$. It follows that the hitting probability vanishes only if $a= e^{-\i(\alpha+\pi/2)}$, i.e. that possible non-hitting only occurs for pure Dirichlet boundary conditions. For these we know that $\alpha= \pi/2+\mu/\sqrt{2}$ on the one-side, and $a=e^{\i \mu/\sqrt{2}}$ on the other side. The two expressions for $a$ are thus only compatible for $\alpha=0$ or $\pi$, i.e. for $\mu = \pm\lambda$. For all values of $\mu$ comprised between $-\lambda$ and $\lambda$ the traces hit the upper boundary of the cylinder with non-zero probability. Second, for $\alpha \in [\pi,2\pi]$ the hitting probability is identically zero provided that $\omega_-=0$. This is equivalent to $\Re a = +\sin \alpha$ and hence $a = e^{\i(\alpha-\pi/2)}$. Using $\alpha= \pi/2+\mu/\sqrt{2}$ we see that this is compatible with pure Dirichlet boundary conditions, provided that $\lambda \leq \mu \leq 3\lambda$. If $\mu$ exceeds the upper bound $3\lambda$ then we are back to the first case because of the $4\lambda = 2\pi R$ periodicity of the compactified free field. Notice the coherence with our previous results: whereas for the uncompactified free field the traces could hit the upper boundary only for $\mu \in (-\lambda,\lambda)$, it is possible in the compactified case for $\mu$-Dirichlet boundary conditions as $\mu \in (-\lambda+2\pi n R,\lambda + 2\pi n R)$, $n\in \mathbb Z$, and more generally for any SU$(2)$ boundary condition which is not pure Dirichlet. In particular, for Neumann boundary conditions with arbitrary $\tilde \mu$ we recover the results from section \ref{sec:ndcase}.

\paragraph{Left-passage probability.} In order to compute the left-passage probability make the following observation. Recall that $f(x,p)$ is a solution to the heat equation $\dot f(x,p) = 2 f''(x,p)$. Using its Lie symmetries we construct new solutions: as pointed out earlier the combination $g(x,p) = 4p f'(x,p)+ x f(x,p)$ solves that equation, too. Define $h_2(x,p) = (g(x,p) - g(-x,p))/4\pi$. Obviously, this function solves the heat equation. Moreover, we have $h_2(x=0,p)=0$ and $h_2(x=2\pi, p) = f(x=2\pi,p)$ for all $p>0$. Let us introduce the auxiliary function
\begin{equation*}
  u_2(x,p) = \frac{h_2(x,p)}{f(x,p)}.
\end{equation*}
We have $u_2(0,p)=0$ and $u_2(2\pi,p)=1$ for all $p>0$. As for the hitting probability we will have to make a distinction between $0\leq \alpha\leq \pi$ and $\pi \leq \alpha \leq 2\pi$. Let us concentrate on the first case. Then for $x\in (0,2\pi)$ we find $\lim_{p\to 0^+} u_2(x,p)= \alpha/(2\pi\omega_+)$ irrespectively of the precise value taken by $x$. Thus the stochastic process $M^{(2)}_t = u_2(Y_t,p-t)$ which is a local martingale for $Y_t$ defined through \eqref{eqn:yprocess}. It is bounded for $0\leq t \leq \tau$ by construction and hence a proper martingale. At $t=\tau$ we can write it as a sum of projectors
\begin{equation*}
  M^{(2)}_\tau = \bm 1_{Y_\tau = 2\pi}+ \left(\frac{\alpha}{2\pi \omega_+}\right)\, \bm 1_{Y_\tau \neq (0,2\pi)}.
\end{equation*}
Thus we find the left-passage probability $\beta(x,p) = \mathsf{E}[M_\tau^{(2)}-\alpha M^{(1)}_\tau/(2\pi \omega_+)] = u_2(x,p)-\alpha u_1(x,p)/(2\pi \omega_+)$, or more explicitly:
\begin{equation}
\beta(x,p) = \frac{\sum_{n\in \mathbb Z}n\left(e^{-{(x-2\alpha-4\pi n)^2}/{8p}}+e^{-{(x+2\alpha-4\pi n)^2}/{8p}}\right)}{\sum_{n\in \mathbb Z} \left(\omega_+\,e^{ -{(x-2\alpha-4\pi n)^2}/{8p}}+ \omega_-\,e^{ -{(x+2\alpha-4\pi n)^2}/{8p}}\right)\label{eqn:su2lp}
}.
\end{equation}
As before, the right-passage probability $\alpha(x,p)$ follows from the sum rule. For $\pi \leq \alpha \leq 2\pi$ the derivation is similar, and the result for $\beta(x,p)$ remains the same.

\section{Conclusion}
\label{sec:conclusion}

In conclusion we have studied different variants of chordal SLE$_4$ on a doubly connected domain. The main outcome were the passage probabilities from the solution of first-exit problems of Brownian bridges and related processes from an interval. These generalise the well-known formula by Schramm for $\kappa=4$. Moreover, using the boundary states of the compactified free field at the self-dual radius we have introduced SLE$_4$ variants parametrised by elements of SU$(2)$, and for which we can continuously interpolate from Dirichlet to Neumann boundary conditions on a second boundary component.

The two values $\kappa=2$ (see \cite{hagendorf:09,phdhagendorf}) and $\kappa=4$, which turn out to be amenable to explicit calculations, are part of the special series $\kappa=2,8/3,4,6,8$ (maybe also $\kappa=3$) where SLE$_\kappa$ enjoys special properties, and which correspond to particularly simple conformal field theories. While it seems difficult to solve the passage problem on doubly connected domains for general $\kappa$ -- given the complexity of the involved partial differential equations -- it would be nice to see whether other values in the special series are amenable to explicit calculations. In our opinion this would be interesting since, as shown in this article, it allows to study how the SLE traces react to different conformal boundary conditions imposed on the second boundary component.

Finally, for $\kappa=4$ it would be interesting to understand whether there is a relation of the SLE$_4$ variants based on the compactified free field at the self-dual radius and some lattice model that admits suitable \textit{lattice} boundary conditions that converge to the SU$(2)$ boundary conditions in the scaling limit.

\subsection*{Acknowledgements}
C.H. would like to thank Costas Bachas, Paul Fendley, Chris{\-}toph Keller, and Pierre Le Doussal for stimulating discussions. Moreover, he thanks the Laboratoire de Physique Th\'eorique de l'ENS where this work was started. The research of C.H. is supported by the NSF grant DMR/MSPA-0704666. The research of D.B. and M.B. is supported by ANR-06-BLAN-0058.

\appendix

\section{Current one-point function and Lie derivatives of the partition function}
\label{app:lie}

Consider a real-valued function $f:G^n \to \mathbb R$ where $G$ is some Lie group. Choose some $X\in \mathfrak g$ from its Lie algebra $\mathfrak g$. We use the Lie derivative $\mathcal L_k^X$ defined as
\begin{equation*}
   (\mathcal L^X_k f)(g_1,\dots,g_n) = \lim_{\epsilon\to 0}\frac{f(g_1,\dots,e^{\i \epsilon X}g_k,\dots,g_n)-f(g_1,\dots,g_n)}{\epsilon}.
\end{equation*}
Here we concentrate on $G = \mathrm{SU}(2)$. For the Lie algebra $\mathfrak{su}(2)$ we choose the spin basis. In the fundamental representation its generators are given by $t^a=\sigma^a/2, \, a=1,2,3$ where $\sigma^a$ are the Pauli matrices.
The partition function $\mathcal A(g_1,g_2)$ depends only on $g_1^{-1}g_2$ and therefore $(\mathcal L_1^X+\mathcal L_2^X)\mathcal A (g_1,g_2)=0$. Let us therefore concentrate on the derivative with respect to $g_1$, which we evaluate for the choice \eqref{eqn:gmatrices}. In fact, it is sufficient to consider the variation of the angle $\alpha$ as $g_1 \to e^{\i \epsilon X} g_1$: if we write $X = \sum_{a=1}^3 X_a t^a$ then we find to first order in $\epsilon$:
\begin{equation*}
  \alpha \to \alpha + \frac{\Re \left[(X_1+\i X_2)b- a X_3\right]}{2\sin \alpha}\,\epsilon+\dots
\end{equation*}
Therefore, the Lie derivative with respect to some general $X = \sum_a X_a t^a$ in $\mathfrak{su}(2)$ is given by
\begin{align}
  \mathcal L^X_1 \mathcal A(g_1,g_2) &=\frac{\Re \left[(X_1+\i X_2)b- a X_3\right]}{2\sin \alpha} \frac{\partial \mathcal A(g_1,g_2)}{\partial \alpha}\nonumber \\
  &=-\frac{\Re \left[(X_1+\i X_2)b- a X_3\right]}{2\sqrt{2}\,\eta(\i p/\pi)\,\sin \alpha}\sum_{n\in \mathbb Z} n \,e^{n^2p/2}\,\sin n\alpha.
  \label{eqn:liepartfunc}
\end{align}
The Lie derivatives $\mathcal L^a\mathcal A$ with respect to the base vectors $t^a$ are found upon specifying $X_b = \delta_{ab}$.

In order to relate this to the one-point function for the current $J^a(z)$ it is convenient to map the cylinder $\mathbb T_p$ back to the annulus $\mathbb A_p$, and use radial quantisation. The conformal transformation from $\mathbb T_p$ to $\mathbb A_p$ is given by $w = e^{\i z}$, and we thus have $\langle J^a(z)\rangle_{\mathbb T_p} = \i w \langle J^a(w)\rangle_{\mathbb A_p}$. On $\mathbb A_p$ the correlation function of any collection of local operators is given by
\begin{equation*}
  \langle \mathcal O \,\rangle_{\mathbb A_p} = \frac{\langle g_1|\mathcal O \,e^{-p(L_0+\bar L_0-1/12)}|g_2\rangle}{\mathcal A}.
\end{equation*}
From this definition and the gluing conditions \eqref{eqn:genbs} it is easy to see that $\langle J^a(w)\rangle_{\mathbb A_p} = w^{-1}\langle J_0^a\rangle_{\mathbb A_p}$. In order to evaluate the insertion of a current zero mode recall that any boundary state $|g\rangle$ may be obtained by the action of $g\cdot \iota$ on a Cardy identity state, and that in some representation $\lambda$ of $\mathfrak{su}(2)$ the zero modes $J^a_0$ act like $-t^a_\lambda$ where $t^a_\lambda$ are the generators of the Lie algebra in that particular representation \cite{difrancesco:97}. Therefore we find
\begin{equation*}
  \mathcal A \langle J^a_0\rangle_{\mathbb A_p} = -\i \mathcal L_1^a \mathcal A = \i \mathcal L_2^a \mathcal A.
\end{equation*}
It follows that $\mathcal A\langle J^a(z)\rangle_{\mathbb T_p} = \mathcal L^a_1 \mathcal A$ as claimed in the text.


\end{document}